%% file: main.tex
\documentclass[fullpage]{article}

\usepackage{xspace}
\usepackage{tikz}
\usepackage{bm}
\usepackage{amsmath}
\usepackage{amsthm}
\usepackage{amssymb}
\usepackage{mathtools}
\usepackage{mathrsfs}
\usepackage{hyperref}
\usepackage{url}
\usepackage[margin=1.25in]{geometry}

\newtheorem{thm}{Theorem}[section]
\newtheorem{theorem}[thm]{Theorem}

\newtheorem{definition}[thm]{Definition}

\newtheorem{lemma}[thm]{Lemma}

\newtheorem{conjecture}[thm]{Conjecture}

\input{macros}

\title{Undefinability of Approximation of 2-to-2 Games}

\author{Anuj {Dawar} \thanks{Funded by UK Research and Innovation (UKRI) under the UK government’s Horizon Europe funding guarantee: grant number EP/X028259/1.}
and B\'alint {Moln\'ar}  \\
Department of Computer Science and Technology, University of
Cambridge, UK  \\
\texttt{anuj.dawar@cl.cam.ac.uk, bm589@cam.ac.uk}}

\begin{document}

\maketitle

\begin{abstract}
Recent work by Atserias and Dawar~\cite{atserias2019definable} and Tucker-Foltz~\cite{TF24} has established undefinability results in fixed-point logic with counting ($\FPC$) corresponding to many classical complexity results from the hardness of approximation.  In this line of work,  $\NP$-hardness results are turned into unconditional $\FPC$ undefinability results.  We extend this work by showing the $\FPC$ undefinability of any constant factor approximation of weighted $2$-to-$2$ games, based on the $\NP$-hardness results of Khot, Minzer and Safra.  Our result shows that the completely satisfiable $2$-to-$2$ games are not $\FPC$-separable from those that are not $\epsilon$-satisfiable, for arbitrarily small $\epsilon$.  The perfect completeness of our inseparability is an improvement on the complexity result, as the $\NP$-hardness of such a separation is still only conjectured.  This perfect completeness enables us to show the $\FPC$ undefinability of other problems whose NP-hardness is conjectured.  In particular, we are able to show that no $\FPC$ formula can separate the $3$-colourable graphs from those that are not $t$-colourable, for any constant $t$.

\end{abstract}

\section{Introduction}
\label{sec:intro}

\input{introduction.tex}

\section{Preliminaries}
\label{sec:prelim}

\input{preliminaries.tex}

\section{The Reduction}

\label{sec:reduction}
\input{reduction.tex}

\section{Definability}
\label{sec:definable}
\input{definability.tex}

\section{Consequences}
\label{sec:conseq}

\input{consequences.tex}

\section{Conclusion}
\label{sec:conclude}
\input{conclusion.tex}

%\input{conclusion.tex}

%%
%% Bibliography
%%

%% Please use bibtex, 

\bibliography{refs}

\end{document}

%% file: macros.tex
\newcommand{\logic}[1]{\ensuremath\mathrm{#1}\xspace}
\newcommand{\FO}{\logic{FO}}

\newcommand{\FPC}{\logic{FPC}}

\newcommand{\cclass}[1]{\ensuremath\mathrm{#1}\xspace}
\newcommand{\NP}{\cclass{NP}\xspace}
\newcommand{\PT}{\cclass{P}\xspace}

\newcommand{\prob}[1]{\ensuremath\textsc{#1}\xspace}

\newcommand{\FF}[1]{\ensuremath{\mathbb{F}_{#1}}\xspace}

\newcommand{\struct}[1]{\ensuremath\mathbb{#1}}
\newcommand{\Aa}{\struct{A}}
\newcommand{\Bb}{\struct{B}}

\newcommand{\class}[1]{\ensuremath\mathscr{#1}}

\newcommand{\ra}{\rightarrow}

\newcommand{\nats}{\mathbb{N}}
\newcommand{\vect}[1]{\overline{#1}}

\newcommand{\UG}{\mathrm{UG}}
\newcommand{\Gap}{\mathrm{Gap}}

\newcommand{\str}[1]{\mathbb{#1}}

%% file: introduction.tex
The study of the hardness of approximation of $\NP$-optimization problems began in earnest with the PCP theorem in the 1990s.  This theorem showed that for many problems (such as~$\prob{MAX 3SAT}$), where there are polynomial-time algorithms that can approximate the optimum solution within a constant factor, there is nonetheless a constant $c$ such that no efficient algorithm can approximate the optimum value within a factor $c$ unless $\PT=\NP$.  Indeed, H{\aa}stad~\cite{Hastad} established tight bounds for $\prob{MAX 3SAT}$: there is a trivial algorithm that achieves an $\frac{8}{7}$ approximation, but none that achieves an $\frac{8}{7}-\epsilon$ approximation for any $\epsilon$, unless $\PT=\NP$.  Such tight bounds are known for many $\NP$-optimization problems, while for others there is a gap in the approximation ratio between the best known algorithm and the strongest known lower bound.  An important problem in the latter category is the \emph{minimum vertex cover} problem, where the best known polynomial-time algorithms yield an approximation ratio of $2$, while the strongest proved lower bound is $\sqrt{2}$.

Perhaps the most important open question in the field of the hardness of approximation is the \emph{unique games conjecture} of Khot.  This states that for any $\epsilon, \delta > 0$, there is a set of labels $\Sigma$ such that it is $\NP$-hard to separate the $(1-\epsilon)$-satisfiable instances of $\Sigma$-unique games (the precise definitions follow below) from those that are not even $\delta$-satisfiable.  The strongest result obtained so far in this direction shows that there is a  $\Sigma$ for which it is $\NP$-hard to separate the $(\frac{1}{2}-\epsilon)$-satisfiable instances from the $\delta$-unsatisfiable ones.  This result is a consequence of the $2$-to-$2$ theorem due to Khot, Minzer and Safra~\cite{Khot16, Khot2016, Khot2018}.

The hardness of approximation has also been studied in recent years in the context of logical definability.  In particular, Atserias and Dawar~\cite{atserias2019definable} showed that many of the $\NP$-hardness results can be recast as \emph{unconditional} undefinability results in \emph{fixed-point logic with counting} ($\FPC$).  For example, there is an  $\FPC$ formula which yields an  $\frac{8}{7}$ approximation of the value of a ~$\prob{MAX 3SAT}$ instance and there is \emph{provably} no formula that  yields an  $\frac{8}{7}-\epsilon$ approximation for any $\epsilon > 0$.  Recall that $\FPC$ is a logic whose expressive power is contained within the complexity class $\PT$ and which has been characterized as a natural \emph{symmetric} fragment of that class~\cite{Anderson2017}.  Tucker-Foltz~\cite{TF24} established the first definability gap in $\FPC$ of unique games, by showing that no formula can distinguish the $\frac{1}{2}$-satisfiable instances from those that are not $(\frac{1}{3}+\delta)$-satisfiable and also showed that no constant factor approximation is $\FPC$ definable.

In the present paper, we consider the $\FPC$ definability of $2$-to-$2$ games.  The hardness of approximating the optimum value of such games was established through a series of results by Khot, Minzer and Safra~\cite{Khot16, Khot2018, Khot2016}.  At the core of their proof is a reduction from the problem~$\prob{MAX 3XOR}$ of maximizing the number of satisfied clauses in a~$\prob{3XOR}$ instance.  We show that the reductions used can be formulated, with some modification, as first-order definable reductions.  As a consequence, we obtain the result that the completely satisfiable instances of  $2$-to-$2$ games cannot be separated by an $\FPC$ formula from those that are no more than $\delta$-satisfiable, for arbitrarily small $\delta$.  This $(1,\delta)$ separation is stronger (in terms of approximation ratios) than the known $(1-\epsilon,\delta)$ $\NP$-hardness result due to the fact that the $\FPC$ undefinability of approximating~$\prob{MAX 3XOR}$  was proved with \emph{perfect completeness} in~\cite{atserias2019definable} (that is to say, with completeness parameter $1$).  A corollary of our result is the $\FPC$ undefinability of a $(\frac{1}{2},\delta)$ separation for unique games.  This improves, again in terms of the approximation ratios, the gap obtained by Tucker-Foltz, though it should be noted that the latter gap is for a more restriced class of games.

A more striking consequence of our result is that no $\FPC$ sentence can separate the class of $3$-colourable graphs from those that are not even $t$-colourable for any constant $t \geq 3$.  The $\NP$-hardness of such a separation has only been proved for $t$ at most $5$, though it is conjectured for larger values.  Indeed, this is a central open problem in the rapidly growing study of \emph{promise constraint satisfaction problems} (PCSP, see~\cite{BBKO21}).

The result on graph colouring should be compared with a recent result of Atserias and Dalmau~\cite{AtseriasDalmau22} which shows that the promise graph colouring problem cannot be solved by a local consistency algorithm.  In particular, this implies that for any constant $t$ the $3$-colourable graphs cannot be separated from those that are not $t$-colourable by a class (whose complement is) definable in Datalog.  Since Datalog programs can be translated into sentences of $\FPC$, our Theorem~\ref{colourability} can be seen as strengthening their result.  It is worth examining this relationship more closely.  It is known, from results of~\cite{ATSERIAS20091666} and~\cite{BartoK}, that every class of bounded counting width (and therefore, in particular, any $\FPC$ definable class) that is the complement of a fixed-template constraint satisfaction problem (CSP) is already definable in Datalog.  Hence, we can conclude from the result of Atserias and Dalmau that no $\FPC$ definable CSP separates the $3$-colourable graphs from the non-$t$-colourable ones.  However, since it is conceivable that a separating class for these two CSP is $\FPC$ definable but not itself a CSP, our result is still a strengthening.
Indeed, in the context of PCSP, the results of~\cite{AtseriasDalmau22} imply that $\FPC$ can define more than Datalog can.  In particular they give an example of a PCSP that is not solvable by local consistency methods but is solvable by a Sherali-Adams lift of the linear programming formulation.  Since we know~\cite{DawarWangCSL} that such lifts of linear programming formulations of CSP are obtained as first-order interpretations and that the solvability of linear programs is expressible in $\FPC$~\cite{AndersonDH15}, it follows that $\FPC$ is strictly more powerful than Datalog for PCSP separations.  Thus, our result on graph colouring is indeed a strengthening.

In Section~\ref{sec:prelim} we introduce the problems, notation and provide background definitions.  An outline of the steps involved in the reduction of Khot, Minzer and Safra is given in Section~\ref{sec:reduction}.  The proof that the reductions involved are definable as first-order interpretations is given in Section~\ref{sec:definable} and certain consequences derived in Section~\ref{sec:conseq}.

%% file: preliminaries.tex
\subsection{Hardness of Approximation in Optimization}

We are interested in $\NP$-hard optimization problems.  A standard example is the problem \\$\prob{MAX 3SAT}$, where the aim is to find, given a  formula in $\prob{3CNF}$, an assignment of values to its variables that maximizes the number of clauses satisfied.  Formally, consider a function problem $M$, which associates with every possible input instance $I$ a value $M(I)$.  In our example, $\prob{MAX 3SAT}$ maps a formula $\phi$ to the maximum number $m$ of clauses of $\phi$ that can be simultaneously satisfied.  While, in practice, we might be interested in finding an assignment that achieves this maximum, for the purpose of proving hardness, it suffices to show that it is hard to compute the number $m$.  When finding $M(I)$ is hard, we may wish to approximate it, and we say that an algorithm computes a $C$-approximation (for a real number $C > 1$) of $M$ if it produces a number $M'(I)$ with the guarantee that $M'(I) \leq M(I) \leq C\cdot M'(I)$.

For the sake of uniformity, we consider function problems that take values in $[0,1]$.  Thus, $\prob{MAX 3SAT}$ assigns to a $\prob{3CNF}$ formula $\phi$ the maximum fraction of the clauses of $\phi$ that can be simultaneously satisfied.  For $\prob{MAX 3SAT}$, it is known that, unless $\PT = \NP$, there is no polynomial-time algorithm that gives a $C$-approximation for any $C < 8/7$.  Such hardness of approximation results are usually proved by means of a \emph{hardness of separation}, which allows us to frame this in terms of the hardness of decision problems.

Formally, let $A$ and $B$ be two sets (i.e.\ decision problems) with $A \cap B = \emptyset$.  We say that $A$ and $B$ are $\NP$-hard to separate, if \emph{every} set $C$ with $A \subseteq C \subseteq \overline{B}$ is $\NP$-hard with respect to Turing reductions, where $\overline{B}$ denotes the complement of $B$ (This definition is symmetric in A and B).  For a function problem $M$, and a constant $c \in [0,1]$, denote by $c$-$M$ the set $\{I \mid M(I) \geq c \}$.  Then, for constants $c$ and $s$ with $0\leq s < c \leq 1$, we say that the \emph{gap problem} $\text{Gap}M(c, s)$ is $\NP$-hard if it is $\NP$-hard to separate the sets $c$-$M$ and $\overline{s\text{-}M}$.  This implies, in particular, that unless $\PT=\NP$, there is no polynomial-time algorithm giving a $\frac{c}{s}$-approximation of $M$.  The value $c$ in $\text{Gap}M(c, s)$ is called the \emph{completeness parameter} and $s$ the \emph{soundness parameter}

The first hardness of approximation results come from the PCP theorem~\cite{PCP98.2, PCP96, PCP98}: one of its direct consequences is the $\NP$-hardness of $\Gap\prob{3SAT}(1,\eta)$ for some constant $\eta$ strictly less than $1$.  H\r{a}stad~\cite{Hastad} obtained an optimum inapproximability result for $\prob{MAX 3SAT}$.  Namely, he showed that $\Gap\prob{3SAT}(1,\frac{7}{8}+\epsilon)$ is $\NP$-hard for arbitrarily small $\epsilon$.  This is optimal since every 3CNF is $\frac{7}{8}$-satisfiable.  Similarly, he also showed that $\Gap\prob{3XOR}(1-\epsilon,\frac{1}{2}+\epsilon)$ is $\NP$-hard for arbitrarily small $\epsilon$.  Again, this is optimal.  Here, $\prob{3XOR}$ is the problem where we are given a Boolean formula as a conjunction of clauses, each of which is the XOR of three literals and we aim to maximize the number of satisfied clauses.  Note that the completeness parameter must be strictly less than $1$, since the problem of determining whether such a formula is satisfiable or not is polynomial-time decidable.  Thus $1$-$\prob{3XOR}$ can be separated in polynomial time from $\overline{(1-\epsilon)\text{-}\prob{3XOR}}$ for any $\epsilon > 0$.

\paragraph*{Reductions}
A common way of deriving further hardness of approximation results is via gap-reductions:  given function problems $A$ and $B$, a polynomial-time computable function $f$ taking instances of $A$ to instances of $B$ is a \emph{reduction} from $\mathrm{Gap}A(c,s)$ to $\mathrm{Gap}B(c',s')$ if for all instances $I$ of $A$
\begin{itemize}
    \item \textbf{Completeness:} if $A(I)\geq c$, then $B(f(I))\geq c'$.
    \item \textbf{Soundness:} if $A(I)\leq s$, then $B(f(I)) \leq s'$.
\end{itemize}
It is easily seen that, if such a reduction exists and $\mathrm{Gap}A(c,s)$ is $\NP$-hard, then so is $\mathrm{Gap}B(c',s')$.

\subsection{Label Cover Games}

Versions of \emph{label cover} problems are ubiquitous in the study of hardness of approximation (see~\cite{DinurS04}).  A particularly important case are the \emph{unique games} of Khot~\cite{KHOT2002}, defined below.  To arrive at the definition, we first introduce some terminology.  For positive integers $d$ and $e$, a relation $R \subseteq U \times V$ is said to be $d$-to-$e$ if it relates each element of $U$ to exactly $d$ elements of $V$ and each element of $V$ to exactly $e$ elements of $U$.

\begin{definition}[$d$-to-$d$ games]\label{def:d-to-d}
  A \emph{$d$-to-$d$ game} is a tuple $(G, \Sigma, \Phi)$, where $G=(V,E)$ is a multi-graph\footnote{That is to say, there may be multiple edges between the same pair of vertices.  In the sequel we refer simply to graphs to mean multi-graphs.}, $\Sigma$ is a finite alphabet and $\Phi: E \rightarrow \mathcal{P}(\Sigma^2)$ assigns to each edge $e \in E$ a $d$-to-$d$ binary relation.

  A \emph{colouring} $\chi: V \rightarrow \Sigma$ \emph{satisfies} an edge $e=(u,v)$ if $(\chi(u), \chi(v)) \in \Phi(e)$.

  The \emph{value} of the game $(G, \Sigma, \Phi)$ is the maximum over all colourings of the proportion of edges in $E$ that are satisfied.

Say that a game $(G, \Sigma, \Phi)$ is \emph{edge consistent} if whenever $e_1$ and $e_2$ are distinct edges on the same pair $(u,v)$ of vertices, we have $\Phi(e_1) = \Phi(e_2)$.

  Say that a game $(G, \Sigma, \Phi)$ is \emph{edge distinct} if whenever $e_1$ and $e_2$ are distinct edges on the same pair $(u,v)$ of vertices, we have $\Phi(e_1) \neq \Phi(e_2)$.

  A game $(G, \Sigma, \Phi)$ is \emph{simple} if it is both edge consistent and edge distinct. In other words, there is at most one edge between any pair of vertices.
\end{definition}
Most of the games we consider with multiple edges are edge consistent.  We mention edge distinct and simple games mainly to constrast with other results in the literature.  For an edge consistent game $(G, \Sigma, \Phi)$, and an edge $e$ between vertices $u$ and $v$, we sometimes write $\Phi(u,v)$ instead of $\Phi(e)$.  No ambiguity arises since by definition of edge consistency, the value of $\Phi$ is the same on all edges between $u$ and $v$.

In this paper, we are particularly interested in $2$-to-$2$ games and $1$-to-$1$ games, the latter also being known as  \emph{Unique Games}.  We write $\UG_q$ for the function problem of determining the value of an instance of unique games with an alphabet of size $q$.  We can then state Khot's unique games conjecture.
\begin{conjecture}[Unique Games Conjecture (UGC)~\cite{KHOT2002}]\label{conj:ugc}
    For any $\delta,\epsilon$ with $0 < \delta < 1-\epsilon < 1$, there exists a positive integer $q$ so that $\Gap\UG_q(1-\epsilon,\delta)$ is NP-Hard.
  \end{conjecture}

  The significance of the conjecture is that it has been shown that many optimal hardness of approximation results follow from it, including $\mathrm{Max\, Cut}$ and $\mathrm{Vertex\, Cover}$~\cite{KHOT2010, Raghavendra2008, KHOT2002}.

  The best known hardness result for unique games, towards proving Conjecture~\ref{conj:ugc} is that $\mathrm{GapUG}_q(\frac{1}{2}-\epsilon,\delta)$ is $\NP$-Hard for arbitrarily small $\delta$ and $\epsilon$.  This is obtained as a consequence of the hardness of $2$-to-$2$ games established by Khot, Minzer and~Safra, which we return to in Section~\ref{sec:reduction}.
  \begin{theorem}[Khot-Minzer-Safra]\label{2-to-2-imperfect}
    For any $\delta,\epsilon$ with $0 < \delta < 1-\epsilon < 1$, there exists a positive integer $q$ so that $\mathrm{Gap2to2}_q(1-\epsilon,\delta)$ is NP-Hard.
  \end{theorem}
It is conjectured that Theorem~\ref{2-to-2-imperfect} can be strengthened to make the completeness parameter $1$, but this remains unproved.
  
 We are also concerned with \emph{weighted} $2$-to-$2$ and $1$-to-$1$ games, attaching a weight to each constraint.
  \begin{definition}[Weighted $d$-to-$d$ games] \label{def:2-to-2}
    A \emph{weighted $d$-to-$d$ game} is a tuple $(G, \Sigma, \Phi,w)$, where $(G,\Sigma,\Phi)$ is a $d$-to-$d$ game and $w: E(G) \ra \mathbb{R}^+$ is a function assigning a positive real weight to each edge.

    Let $\mathrm{tot} = \sum_{e \in E(G)} w(e)$ be the total weight.
      The \emph{value} of the game $(G, \Sigma, \Phi, w)$ is the maximum over all colourings $\chi: V \ra \Sigma$ of the fraction $\sum_{e \in S_{\chi}} w(e) /\mathrm{tot}$, where $S_{\chi}$ denotes the set of edges $e = (u,v)$ for which $(\chi(u),\chi(v)) \in \Phi(e)$.
\end{definition}

We write $\mathcal{WG}_{2:2;q}$ to denote the class of weighted $2$-to-$2$ games with $q$ labels and $\prob{Weight2to2}_q$ to denote the function taking such a game to its value.  Similarly, we write $\prob{UG}_q$ and $\prob{WeightUG}_q$ for the functions giving the values of unique games and weighted unique games with $q$ labels respectively.

Note that if in a game $(G, \Sigma, \Phi,w)$, all weights $w(e)$ are integers, than this is equivalent to an \emph{unweighted} game where we replace each edge $e$ of $G$ with $w(e)$ distinct edges, each labelled with the constraint $\Phi(e)$.  Importantly, this unweighted game is \emph{edge consistent} but not \emph{simple}.
  
  \subsection{Undefinability of Approximation}

  We assume the reader is familiar with first-order logic and the basics of finite model theory.  A good introduction is to be found in~\cite{EbbinghausF99}.  Our structures are finite structures in a finite relational vocabulary.  Our main inexpressibility results are stated for \emph{fixed-point logic with counting} ($\FPC$).  We do not need a formal definition here but note that every property definable in $\FPC$ is decidable in polynomial-time and indeed $\FPC$ can be understood as a complexity class defined by \emph{symmetric polynomial-time} computation.  For full definitions, refer to~\cite{Dawar15} and references therein.

  The two properties of $\FPC$ that we do need are that (1) every class of structures definable in $\FPC$ has \emph{bounded counting width}; and (2) that the class of properties definable in $\FPC$ is closed under \emph{first-order interpretations}.  We elaborate on these below.

  For a function problem $M$, and real numbers $c$ and $s$ with $0 \leq s < c \leq 1$, we say that $\text{Gap}M(c, s)$ is undefinable in $\FPC$ if there is no $\FPC$ definable class of structures that separates the sets $c$-$M$ and $\overline{s\text{-}M}$.
  Atserias and Dawar~\cite{atserias2019definable} initiated a study of the $\FPC$ undefinability of approximations, showing that many of the $\NP$-hardness results for gap problems can be reproduced as \emph{unconditional} undefinability results in $\FPC$.  In particular $\Gap\prob{3SAT}(1,\frac{7}{8}+\epsilon)$ is not $\FPC$ definable.  More significantly, they established the following
  \begin{theorem}[Atserias-Dawar~\cite{atserias2019definable}]\label{thm:fpc-xor}
  For any $\epsilon$ with $0 < \epsilon < \frac{1}{2}$, $\Gap\prob{3XOR}(1,\frac{1}{2}+\epsilon)$ is not $\FPC$ definable.    
  \end{theorem}
  Note the completeness parameter of $1$ in the statement, which contrasts with $1 - \epsilon$ in the case of Theorem~\ref{2-to-2-imperfect}.  Perfect completeness cannot be established in the case of $\NP$-hardness because satisfiability of XOR formulas \emph{is} decidable in polynomial-time.  However, it is not definable in $\FPC$ and this allows the stronger result in the context of undefinability.  This is crucial to the application we make of Theorem~\ref{thm:fpc-xor} in Section~\ref{sec:colouring}

  Following up on this work, Tucker-Foltz~\cite{TF24} studied the undefinability of gaps in unique games.  In particular, he established the inapproximability of unique games in $\FPC$ by any constant factor and the $\FPC$-undefinability of  $\Gap\UG_q(\frac{1}{2},\frac{1}{3} + \delta)$ for a suitable value of $q$.

\paragraph*{Counting Width}
  For relational structures $\str{A}$ and $\str{B}$ in the same vocabulary, and a positive integer $k$, $\str{A} \equiv_{C^k} \str{B}$ denotes that the two structures cannot be distinguished by any sentence of first-order logic \emph{with counting} using no more than $k$ distinct variables.  For a class $\mathcal{C}$ of structures, the \emph{counting width} of $\mathcal{C}$ is the function $\nu : \nats \rightarrow \nats$ such that for any $n$, $\nu(n)$ is the least $k$ such that $\mathcal{C}$, restricted to structures with at most $n$ elements is a union of $\equiv_{C^k}$-equivalence classes.   Any class that is definable by a sentence of $\FPC$ has counting width bounded by a constant. Almost all results showing that a class is not definable in $\FPC$ proceed by showing that it, in fact, does not have bounded counting width.  Note that the \emph{Weisfeiler-Leman} dimension of a graph $G$ is defined to be the smallest $k$ such that for any graph $H$ that is not isomorphic to $G$, we have $G \not\equiv_{C^{k+1}} H$.  This is a graph parameter associating a number with a graph in contrast to counting width (as defined in~\cite{DawarWangCSL}) which associates a function with a class $\mathcal{C}$.

  \paragraph*{Interpretations}
  A \emph{first-order interpretation} of a relational vocabulary $\tau$ in a vocabulary $\sigma$ is a sequence of $\sigma$-formulas in first-order logic, which can be seen as mapping $\sigma$-structures to $\tau$-structures.  There are many variations of the precise definition in the literature.  We use the version defined in~\cite{atserias2019definable} and include the definition here for completeness. 

  Consider two vocabularies~$\sigma$ and~$\tau$.  A \emph{$d$-ary
  ~$\FO$-interpretation of~$\tau$ in~$\sigma$} is a sequence of
first-order formulas in vocabulary~$\sigma$ consisting of: (i) a
formula~$\delta(\vect{x})$; (ii) a formula~$\varepsilon(\vect{x},
\vect{y})$; (iii) for each relation symbol~$R \in \tau$ of arity~$k$,
a formula~$\phi_R(\vect{x}_1, \dots, \vect{x}_k)$; and (iv) for each
constant symbol~$c \in \tau$, a formula~$\gamma_c(\vect{x})$, where
each~$\vect{x}$,~$\vect{y}$ or~$\vect{x}_i$ is a~$d$-tuple of
variables.  We call~$d$ the \emph{dimension} of the interpretation.    We say
that an interpretation~$\Theta$ associates a~$\tau$-structure~$\Bb$ to
a~$\sigma$-structure~$\Aa$ if there is a map~$h$ from~$\{ \vect{a} \in
A^d \mid \Aa \models \delta[\vect{a}] \}$ to the universe~$B$ of~$\Bb$
such that: (i)~$h$ is surjective onto~$B$; (ii)~$h(\vect{a}_1) =
h(\vect{a}_2)$ if, and only if,~$\Aa\models \varepsilon[\vect{a}_1,
  \vect{a}_2]$; (iii)~$R^{\Bb}(h(\vect{a}_1), \dots, h(\vect{a}_k))$
if, and only if,~$\Aa \models \phi_R[\vect{a}_1, \dots, \vect{a}_k]$;
and (iv)~$h(\vect{a}) = c^{\Bb}$ if, and only if,~$\Aa \models
\gamma_c[\vect{a}]$.  Note that an interpretation~$\Theta$ associates
a~$\tau$-structure with~$\Aa$ only if~$\varepsilon$ defines an
equivalence relation on~$A^d$ that is a congruence with respect to the
relations defined by the formulae~$\phi_R$ and~$\gamma_c$. In such
cases, however,~$\Bb$ is uniquely defined up to isomorphism and we
write~$\Theta(\Aa) = \Bb$.  For a class of structures~$\class{C}$ and an interpretation~$\Theta$,
we write~$\Theta(\class{C})$ to denote the class~$\{\Theta(\struct{A})
\mid \struct{A} \in \class{C}\}$.  We mainly use interpretations to
define reductions between classes of structures. 
  
  Given a function problem $A$ whose instances are $\sigma$-structures and a function problem $B$ whose instances are $\tau$-structures, an interpretation $\Theta$ of $\tau$ in $\sigma$ is a $\mathrm{Gap}A(c,s)$ to $\mathrm{Gap}B(c',s')$ reduction if $A(\str{A})\geq c$ implies $B(\Theta(\str{A}))\geq c'$ and $A(\str{A})\leq s$ implies $B(\Theta(\str{A})) \leq s'$.  Definability in $\FPC$ and the property of having bounded counting width are both closed under first-order reductions.  That is to say, if $\mathrm{Gap}B(c',s')$ is $\FPC$-definable and there is a first-order reduction of $\mathrm{Gap}A(c,s)$ to $\mathrm{Gap}B(c',s')$, then $\mathrm{Gap}A(c,s)$ is $\FPC$-definable as well.

%% file: reduction.tex
The proof of Theorem~\ref{2-to-2-imperfect} was completed in 2018 and remains to this day the most significant advance towards establishing the Unique Games Conjecture since the latter was formulated by Khot in~\cite{KHOT2002}.  The proof proceeds by a reduction from $\Gap\prob{3XOR}(1-\epsilon,\frac{1}{2}+\delta)$ and was presented in a series of papers~\cite{Khot16, Khot2018, Khot2016}.  The main difficulty lies in proving the combinatorial conditions that the soundness analysis relies on.  The full reduction and proof of correctness can be found in~\cite[Chapter~3]{Minzer18}.

Our aim in the present paper is to show that the reduction constructed has two crucial properties.  First, it preserves perfect completeness and thus can be seen as a reduction from $\Gap\prob{3XOR}(1,\frac{1}{2}+\delta)$.  Secondly, with small modifications which do not affect the soundness or completeness analysis, it can be described as a first-order interpretation.  Together these establish the main theorem.
\begin{theorem}\label{FPC-2-to-2}
   For every $\delta$ with $0 < \delta < 1$, there exists $q\in \mathbb{N}^+$ for which $\Gap\prob{2to2}_q(1,\delta)$ is not $\FPC$ definable.
 \end{theorem}

 In proving this, we do not need to reprise the difficult soundness analysis carried out by Khot et al.  Rather we study the actual construction involved in the reduction.  For this purpose, we describe the reduction in some detail in this section, and take up the two issues of perfect completeness and first-order definability in the next.  The reduction as described here produces a \emph{weighted} game.  We also discuss in the next section how the weights may be chosen to be integers and bounded by a polynomial function of the instance size.

 \subsection{Regular 3XOR}

 An instance of $\prob{3XOR}$ can be seen as a system of linear equations over the field $\FF{2}$ with exactly three variables appearing in each equation.  We say that such an instance is \emph{$d$-regular} if every variable appears in at most $d$ equations and no two equations share more than one variable.  It is known that for some constant $\frac{1}{2}<\eta<1$, the $\NP$-hardness of  $\Gap\prob{3XOR}(1-\epsilon, \eta)$ holds even when restricted to \emph{$d$-regular} instances for some fixed value of $d$ (indeed, taking $d = 5$ suffices, see~\cite[Theorem~3.3.1]{Minzer18}).  In Section~\ref{sec:regular} we show that this is also true of the undefinability in $\FPC$ of $\Gap\prob{3XOR}(1,\eta)$
 From now on, we restrict attention to $d$-regular instances for a suitable fixed value of $d$, and we call the resulting function problem $\Gap\prob{Regular3XOR}$. 
 
 This may appear to be an unusual usage of the word ``regular'' since we only have an upper bound on the number of appearances of a variable in the equations and it does not correspond to regularity in hypergraphs. However, this is the  terminology of the papers by Khot, Minzer, and Safra and we use it for consistency.

\subsection{Reducing to Transitive Games}\label{sec:transitive}

In the first step of the reduction, we reduce regular $\prob{3XOR}$ instances to label cover games with a mixture of $2$-to-$2$ and $1$-to-$1$ constraints, with an additional transitivity requirement.  We formally define these below.

\begin{definition}[Transitive 2-to-2 games]
  An edge consistent game $(G, \Sigma, \Phi)$ is a \emph{transitive 2-to-2 game} if $\Phi: E \rightarrow \mathcal{P}(\Sigma^2)$ assigns to each edge $e$ either a $2$-to-$2$ or a $1$-to-$1$ relation and whenever $\Phi(u,v)$ is $1$-to-$1$, then for any edge $(v,w)$, $\Phi(u,w)$ is the composition of $\Phi(u,v)$ and $\Phi(v,w)$.
\end{definition}
Note that the condition on composition only applies when $\Phi(u,v)$ is $1$-to-$1$, but $\Phi(v,w)$ may be $1$-to-$1$ or $2$-to-$2$, and this determines whether $\Phi(u,w)$ is $1$-to-$1$ or $2$-to-$2$.    

Now, fix an instance $I$ of $\prob{GapRegular3XOR}$, with $X$ being the set of variables that appear in $I$ and $E$ the set of equations.  Thus, each equation $e \in E$ is of the form $x + y+ z = b$ for some $b \in \FF{2}$.  We refer to $x,y$ and $z$ as the variables occurring in $e$ and $b$ as the \emph{right-hand side} of $e$.

Fix a positive integer $k$ and let $\mathcal{U} \subseteq E^k$ be the set of $k$-tuples $U$ of equations, satisfying the following properties:
\begin{itemize}
    \item no variable occurs in more than one equation of $U$; and
    \item if variables $x$ and $y$ appear in distinct equations of $U$, there is no equation in $E$ (even outside $U$) in which both $x$ and $y$ occur.
 \end{itemize}

 For $U = (e_1,\ldots,e_k) \in \mathcal{U}$, let  $X_U$ denote the set of variables occuring in equations in $U$ and for $i \in \{1,\ldots,k\}$  let $v_i \in \FF{2}^{X_U}$ denote the vector which has  $1$s in the three coordinates corresponding to the variables occurring in $e_i$ and $0$s everywhere else.  We define the \emph{space of side-conditions} corresponding to $U$ to be $H_U = \mathrm{Span}(v_1, \dots, v_k)$.  Note that, by construction, the vectors $v_1,\ldots,v_k$ are linearly independent and thus form a basis for $H_U$.  We say that a linear function $f: \FF{2}^{X_U} \ra \FF{2}$ satisfies the equations in $U$ if $f(v_i)=b_i$ for all $i$, where $b_i$ is the right-hand side of $e_i$.

 Now, fix a parameter $l$ with $l < k$, and we define $\mathcal{L}_U$ to be the collection of $l$-dimensional subspaces of $\FF{2}^{X_U}$ which are linearly independent of $H_U$.  That is
 \[\mathcal{L}_U = \left\{ L \subseteq \FF{2}^{X_U} \mid \mathrm{dim}(L)=l,\, L \cap H_U  = \{\bm{0}\}\right\}.\]
 The trivial intersection ensures that for any subspace $L \in \mathcal{L}_U$, any linear function $f: L \ra \FF{2}$ can be uniquely extended to one on $L +H_U$\footnote{Here the sum is to be understood as vector space sum, i.e.\ $L+H_U$ is the space spanned by the union of $L$ and $H_U$.} so that $f(v_i) = b_i$ for all $i$. Therefore, the number of linear functions on $L + H_U$ satisfying the equations in $U$ is exactly $2^l$.

 We can now define the reduction $\Theta$ that takes the instance $I$ to a $2$-to-$2$ transitive game $\Theta(I)$. We use parameters $k$ and $l$ throughout the construction, which only depend on $\epsilon$ and $\delta$ (for the purpose of the reduction, they are constants). Their exact dependence on $\epsilon$ and $\delta$ can be found in \cite[Section 3.4.5]{Minzer18}. Notably, the choices of $k$ and $l$ are only independent of the input size, but not independent of each other, and the construction requires $l$ to be significantly smaller than $k$.

 \noindent
 \textbf{Vertices}

 The vertices of $\Theta(I)$ are pairs $(U, L)$, where $U \in \mathcal{U}$ and $L \in \mathcal{L}_U$.

 \noindent
 \textbf{Alphabet}

The alphabet is a set of labels of size $2^l$. As noted above, for each vertex  $(U, L)$, there are exactly $2^l$ linear functions on $L + H_U$ satisfying the equations in $U$.  We fix, for each $(U, L)$, a bijection between the alphabet and this set of linear functions.  Henceforth, we simply treat the functions themselves as labels.

\noindent
\textbf{Constraints}

Given a pair of vertices $u = (U, L)$ and $v = (U', L')$, the constraint $\Phi(u,v)$ is a $1$-to-$1$ relation if
\[\mathrm{dim}(L + H_U + H_{U'}) = \mathrm{dim}(L' + H_U + H_{U'}) = \mathrm{dim}(L + L' + H_U + H_{U'})\] and a $2$-to-$2$ relation if
\[\mathrm{dim}(L + H_U + H_{U'}) = \mathrm{dim}(L' + H_U + H_{U'}) = \mathrm{dim}(L + L' + H_U + H_{U'}) - 1.\]
We relate $f$ to $f': L' +  H_{U'} \ra \FF{2}$ if, and only if, $f$ and $f'$ agree on the shared space $(L + H_U + H_{U'}) \cap  (L' + H_U + H_{U'})$.

It is the case for any pair, that $\mathrm{dim}(L + H_U + H_{U'}) = \mathrm{dim}(L' + H_U + H_{U'})$~\cite[Lemma~4.3]{Khot16}.  Let us call this dimension $D$.  By~\cite[Lemma~4.4]{Khot16}, any linear function $f: L + H_U \rightarrow \mathbb{F}_2$ satisfying the equations of $U$ has a unique extension to  $(L + H_U + H_{U'})$ that also satisfies the equations of $U'$.  Then, it is easily seen that if $\mathrm{dim}(L + L' + H_U + H_{U'}) = D$, then $f$ has exactly one label of $(U',L')$ that it is consistent with, and if $\mathrm{dim}(L + L' + H_U + H_{U'}) = D + 1$, there are exactly two such functions, thanks to the ``free dimension''.  Hence, the constraints are $1$-to-$1$ or $2$-to-$2$ as required.  The transitivity property of these constraints is established in~\cite[Appendix A]{Khot16/2}. A more detailed explanation on this can also be found in ~\cite[Appendix A]{Khot16/2}.

\subsection{The final (weighted) 2-to-2 game}\label{final-weighted-2-to-2-game}

The final step of the reduction is to transform the transitive game constructed in Section~\ref{sec:transitive} into a \emph{weighted} $2$-to-$2$ game, getting rid of the $1$-to-$1$ constraints.  This weighted game is defined as follows.

Recall the transitive $2$-to-$2$ game $\Theta(I)$ constructed in Section~\ref{sec:transitive}.  The transitivity condition guarantees that the vertices of $\Theta(I)$ can be partitioned into cliques $C_1, \dots, C_m$ so that edges in each clique are associated with $1$-to-$1$ constraints.  Moreover, these constraints are consistent in the sense that any colouring of a vertex $V$ in a clique $C$ can be extended in a unique way to a colouring of all vertices in $C$ so that all edge constraints in $C$ are satisfied.  Also, by the transitivity condition, for distinct cliques $C_i$ and $C_j$, either all pairs $(u,v) \in  C_i  \times C_j$ are connected by $2$-to-$2$ constraints or none are.  Furthermore, these $2$-to-$2$ constraints are consistent in the sense that given a clique-consistent colouring for $C_i$ and $C_j$, either all or none of these $2$-to-$2$ constraints are satisfied.

The final (weighted) 2-to-2 instance $I_{2:2}^w$ we construct from $\Theta(I)$ has as vertices the vertices of $\Theta(I)$ and as edges all pairs $(u,v)$ in $\Theta(I)$ where $u$ and $v$ are in distinct cliques.  For each such edge, with $u \in C_i$ and $v \in C_j$, we associate the constraint $\Phi(u,v)$ which is as in $\Theta(I)$.  The weight $w(u,v)$ is the probability assigned to $(u,v)$ by the following sampling process\footnote{Some weights may be $0$. Those edges are dropped from the final construction in Section 4}:
\begin{itemize}
    \item Choose $U \in \mathcal{U}$, uniformly at random.
    \item Choose a random pair $L, L'$ so that $(U,L)$ and $(U,L')$ are connected by a $2$-to-$2$ edge.  Let $C_i$ be the clique containing $(U,L)$ and $C_j$ be the clique containing $(U,L')$ 
    \item Choose uniformly at random a pair of vertices $(u,v) \in C_i \times C_j$. 
\end{itemize}

\subsection{Irregular soundness case}

For the result in Section~\ref{sec:colouring}, we need the $\FPC$-undefinability of a different gap problem based on $2$-to-$2$ games.  Specifically, we define the value of a game to be, not the fraction of constraints that can be satisfied, but the fraction of the vertices formed by the largest set $X$ so that all constraints between nodes in $X$ are satisfied.  Moreover, we relax the notion of colouring to allow vertices to be coloured by multiple colours.

\begin{definition}
  For a $2$-to-$2$ game $((V,E),\Sigma, \Phi)$, a colouring $c: V \rightarrow \binom{\Sigma}{j}$ satisfies a set $X \subseteq V$ if  $\forall e \in E$, such that the vertices of $e$ are in $X$,  $\exists a \in c(u), b \in c(v). (a,b) \in \Phi(e)$.
\end{definition}

That is to say, a $j$-colouring, i.e., one that assigns a set of $j$ colours to each vertex satisfies a set $X$ if each constraint between vertices in $X$ is satisfied by some choice among the colours assigned to the vertices.

\begin{definition}[Irregular Values]
  For constants $j$ and $q$ define the function $\prob{Irreg2to2}_{j,q}$ to take a $2$-to-$2$ game $((V,E),\Sigma, \Phi)$ to the fraction $|X|/|V|$ where $X$ is the largest subset of $V$ that is satisfied by some $j$-colouring $c: V \rightarrow \binom{\Sigma}{j}$ .
\end{definition}

\begin{theorem}[Definable 2-to-2 Games Theorem with irregular soundness]\label{irregular-FPC-2-to-2}
For every $\delta$ with $0 < \delta < 1$ and $j\in \mathbb{N}^+$, there exists $q\in \mathbb{N}^+$ so that $\Gap\prob{Irreg2to2}_{j,q}(1,\delta)$ is not $\FPC$ definable.
\end{theorem}

Note that the value  $\prob{Irreg2to2}_{j,q}((V,E),\Sigma, \Phi)$ is unchanged if we \emph{simplify} the game (i.e.\ remove edge weights and multiplicities of the same constraint between a pair of nodes), therefore Theorem~\ref{irregular-FPC-2-to-2} applies even in restriction to simple, unweighted 2-to-2 games.

It is not hard to see that this is a consequence of the proof of Theorem~\ref{FPC-2-to-2}, and the corresponding claim for $\NP$-hardness appears in e.g.~\cite{Khot16}.  For completeness, we give a short account of the reduction, which combines with the proof of Theorem~\ref{FPC-2-to-2} to yield Theorem~\ref{irregular-FPC-2-to-2}.

\begin{lemma}
  \label{reg-to-irregular}
 For a weighted simple $2$-to-$2$ game $I = ((V,E),\Sigma, \Phi)$ as constructed in Section~\ref{final-weighted-2-to-2-game} with $q = |\Sigma|$, if $\prob{Irreg2to2}_{j,q}((V,E),\Sigma, \Phi) \geq \delta$, then $\prob{Weight2to2}(I) \geq \frac{\gamma(\delta)}{j^2}$ for some $\gamma(\delta) \in \Omega(\delta^2)$.
\end{lemma}
\begin{proof}
     Let $c$ be a $j$-colouring of $V$ that  satisfies a set $X$ with $|X|/|V| \geq \delta$.  By \cite[Remark~3.4.9]{Minzer18},  there is a $\gamma(\delta) \in \Omega(\delta^2)$ (weighted) fraction of the edges $E$ which are satisfied by $c$, in the sense that for each such edge $e$ on vertices $u,v$ there are colours $a$ and $b$ in $c(u)$ and $c(v)$ respectively such that $(a,b) \in \Phi(e)$.  We now construct a standard colouring by a random process.  That is, for each vertex $v \in V$, independently choose a colour $\chi(v)$ from $c(v)$ uniformly at random.  For an edge $e$ on vertices $u,v$, let $\Xi(e)$ be the indicator variable indicating whether $(\chi(u),\chi(v)) \in \Phi(e)$ and let $\Xi$ be the overall value of the colouring $\chi$.     If $(u,v) \in X^2$, the probability that $\chi$ satisfies the constraint $\Phi(e)$ is at least $\frac{1}{j^2}$, as by definition, among the $j^2$  pairs in $c(u) \times c(v)$, at least one satisfies the constraint.  Then, we have,
    \begin{align*}
        \mathbb{E}[\Xi] &= \mathbb{E}[\sum_{e \in E} w(e) \Xi(e)] =  \sum_{e \in E} w(e) \mathbb{E}[\Xi(e)] \\
        &\geq \sum_{e \in E|_X} w(e) \mathbb{E}[\Xi(e)] \geq \sum_{e \in E|_X} w(e) \frac{1}{j^2} \geq \gamma(\delta) \frac{1}{j^2}
    \end{align*}
    where $E|_X$ denotes the subset of $E$ containing those edges both of whose vertices are in $X$.
    Thus, there is a colouring that satisfies at least an $\Omega(\delta^2)$ (weighted) fraction of the constraints.
\end{proof}

\subsection{$2 \leftrightarrow 2$ games} \label{hourglass-variant}
The definition of $2$-to-$2$ games, Definition~\ref{def:2-to-2} only requires each constraint $\Phi(u,.v)$ to be a $2$-to-$2$ relation, meaning that each element on the left is related to exactly two elements on the right and vice versa.  However, the reductions yield games of a more restricted kind, and this will be useful in Section~\ref{sec:colouring}.  Say that a binary relation $R \subseteq A \times B$ is $2 \leftrightarrow 2$ if it is the disjoint union of bipartite graphs $K_{2,2}$.  That is to say, $A$ and $B$ can each be partitioned into sets $A = \bigcup_i A_i$ and  $B = \bigcup_i B_i$ so that each $A_i$ and $B_i$ has exactly two elements and $R = \bigcup_i A_i \times B_i$.

We claim that the reductions in the proof of Theorem~\ref {FPC-2-to-2} yield games in which all constraint relations are $2 \leftrightarrow 2$.  Specifically, given a fixed pair of nodes $(U,L)$ and $(U',L')$, and two different linear functions $f, f': L + H_U \rightarrow \mathbb{F}_2$ so that their unique extension to the domain $L + H_U + H_{U'}$ only differ in their ``free dimension", i.e. they agree in values on $( L + H_U + H_{U'}) \cap (L' + H_U + H_{U'})$, $f$ and $f'$ are related to the same two linear functions on $L' + H_{U'}$ (uniquely extensible to $L' + H_U + H_{U'}$) in $\Phi((U,L),(U',L'))$.  Thus, the constraint relations constructed are $2 \leftrightarrow 2$.

Notice, that any $2 \leftrightarrow 2$ constraint $\Phi$ can be represented by a pair of permutations $\pi_1, \pi_2$, so that $\phi = \{(\pi_1(2\kappa+\iota_1), \pi_2(2\kappa +\iota_2)) \mid 0 \leq \kappa < \frac{q}2, \iota_1, \iota_2 \in \{1,2\} \}$ (for a fixed $\kappa$, $\pi_1(2\kappa), \pi_1(2\kappa+1), \pi_2(2\kappa), \pi_2(2\kappa+1)$ form a $K_{2,2}$ component in the constraint graph), given a fixed (arbitrary) bijection between the label set and $\{1, 2, ..., q\}$.  We call such a constraint $D_{\pi_1,\pi_2}$. Notice that any $2 \leftrightarrow 2$ game must have an even alphabet. Also notice that multiple pairs of permutations can describe the same constraint.

%% file: definability.tex
The aim in this section is to show that the reduction outlined in Section~\ref{sec:reduction} can, with minor modifications, be implemented as a first-order interpretation, preserving perfect completeness.  Thus, it gives a first-order definable reduction from $\Gap\prob{3XOR}(1, \frac{1}{2} + \delta)$ to $\Gap\prob{2to2}_q(1,\delta')$ for a suitable choice of parameters.  Moreover, as we show in Section~\ref{weight approximation}, the weights can be chosen to be integers and bounded by a polynomial in the instance size.  This allows us to reduce them to unweighted, edge consistent games.  This establishes Theorem~\ref{FPC-2-to-2}.

\subsection{Perfect completeness}

To show that the reduction from Section~\ref{sec:reduction} preserves perfect completeness, it suffices to verify that instances of $\prob{3XOR}$ that are satisfiable (i.e.\ have value $1$) are mapped by the reduction to instances of $\mathcal{WG}_{2:2}$ which also have value $1$.

Assume $I$ is a 3XOR instance on a set of variables $X$ that is satisfiable, and let
$s : X \to \mathbb{F}_2$ be an assignment of values to the variables that satisfies it.
Let $f_s : \mathbb{F}_2^{X} \to \mathbb{F}_2$ denote the linear map induced by $s$, i.e.
$f_s(x) = \langle s, x \rangle$, where we identify $s$ with the vector
$(s(x))_{x \in X} \in \mathbb{F}_2^{X}$. Let $I^w_{2:2}$ denote the weighted 2-to-2 game
that $I$ maps to under the reduction. Then, for each vertex $(U, L)$ of $I^w_{2:2}$, the
restriction of $f_s$ to $L + H_U$ is a valid label: it is linear by construction, and since
$s$ satisfies every equation of $I$, we have $f_s(v_i) = b_i$ for each equation $e_i$ of $U$,
so the restriction satisfies the equations in $U$. Moreover, for any constraint between
$(U,L)$ and $(U',L')$, the labels $f_s|_{L + H_U}$ and $f_s|_{L' + H_{U'}}$ agree on the
intersection of their domains, being restrictions of the same global function $f_s$, and
hence satisfy the constraint. Thus this labelling satisfies all constraints and
$\mathrm{Weight2to2}(I^w_{2:2}) = 1$.

\subsection{Vocabularies} \label{vocabularies}

An instance of $\prob{3XOR}$ is defined as a structure over the vocabulary $\tau_{\mathrm{3XOR}} = \langle \mathrm{Eq}_0, \mathrm{Eq}_1 \rangle$ with two ternary relations.  We think of the universe of a $\tau_{\mathrm{3XOR}}$-structure $\str{A}$ as a set of variables.  For $b \in \{0,1\}$, a triple $(x,y,z) \in \mathrm{Eq}_b$ is understood as representing the equation $x + y + z = b$, where addition is modulo $2$.

For each positive integer $q$, we define a vocabulary $\tau_{\text{(T) 2-to-2}_{q}}$ such that structures in this vocabulary represent instances of transitive $2$-to-$2$ games over a label alphabet of size $q$.  Let $S_q$ denote the collection of permutations of $[q] = \{1,\ldots,q\}$.  Note that there is a natural bijective correspondence between $S_q$ and the $1$-to-$1$ relations on $[q]$.  Now, let  $S^{\#2}_q$ denote the set of pairs of permutations $(\pi_1,\pi_2) \in S_q \times S_q$ such that for all $i \in [q]$, $\pi_1(i) \neq \pi_2(i)$.  Then, it is easily seen that each $2$-to-$2$ relation on $[q]$ can be seen (not uniquely) as the union of such a pair of permutations.  In particular, each $i$ is related to exactly the two elements $\pi_1(i)$ and $\pi_2(i)$.  Our vocabulary $\tau_{\text{(T) 2-to-2}_{q}}$ contains a binary relation for each element of $S_q$  and one for each element of $S^{\#2}_q$:
\[\tau_{\text{(T) 2-to-2}_{q}} = \langle (C_\pi)_{\pi \in S_q}, (C_{\pi_1, \pi_2})_{(\pi_1,\pi_2) \in S_q^{\#2}}\rangle.\]
  We write $\mathcal{C}_1$ for the collection of relation symbols $ (C_\pi)_{\pi \in S_q}$ and $\mathcal{C}_2$ for the collection of relation symbols $(C_{\pi_1, \pi_2})_{(\pi_1,\pi_2) \in S_q^{\#2}}$. Note that the vocabulary itself does not enforce the transitivity property.  Also note that each pair $(u,v)$ of vertices can appear in multiple relations.  However, our formulation does not allow for multiple constraints of the same type between the same pair $(u,v)$.  In other words, the games are \emph{edge distinct} in the sense of Definition~\ref{def:d-to-d}.  This suffices because the transitive games constructed in Section~\ref{sec:transitive} are simple.

    For general (that is not necessarily edge-distinct) $2$-to-$2$ games, we adopt a representation in which both vertices and edges are first-class elements.  This allows for multiple edges between a pair of vertices, which may or may not all be associated with the same constraint. Specifically, 
    \[\tau_{\text{(w) 2-to-2}_{q}} = \langle C, (\Phi_{\pi_1, \pi_2})_{(\pi_1,\pi_2) \in S_q^{\#2}}\rangle,\]
      where $C$ is unary, and the relations $\Phi_{\pi_1, \pi_2}$ are all ternary.  In a $\tau_{\text{(w) 2-to-2}_{q}}$-structure $\str{A}$, the universe of $\str{A}$ is the disjoint union of the set $V$ of vertices of  $I^w_{2:2}$, and the set $C$ of constraints, with the unary relation $C$ picking out this set.  For each $(\pi_1,\pi_2) \in S_q^{\#2}$, the relation $\Phi_{\pi_1,\pi_2} \subseteq V^2 \times C$ contains those triples $(u,v,c)$ where $c$ is a constraint linking $u$ and $v$ and $\Phi(c)$ is the $2$-to-$2$ relation associated with the pair $(\pi_1,\pi_2)$.  We assume our structures satisfy the (first-order) axiom that ensures that for each $c$, there is exactly one pair $(u,v)$ and one relation $\Phi_{\pi_1,\pi_2}$ in which the triple $(u,v,c)$ appears.  Note that a structure in this vocabulary can also be understood as a \emph{edge distinct, weighted} game with integer weights.  The weight of a constraint on an edge $(u,v)$ labelled with relation $R$ is simply the number of constraints $c$ for which $(u,v,c)$ is in $R$.

\subsection{Undefinability of Regular 3XOR}\label{sec:regular}
The reduction in Section~\ref{sec:reduction} starts from \emph{regular} instances of $\prob{3XOR}$.  In contrast, the undefinability result in Theorem~\ref{thm:fpc-xor} is stated for general $\prob{3XOR}$.  Thus, we begin by arguing that the proof of Theorem~\ref{thm:fpc-xor} can actually be used to show the $\FPC$ undefinability of $\Gap\prob{Regular3XOR}(1,\eta)$ for some $\eta$ strictly smaller than $1$. We break down the proof into two parts.

\subsubsection*{Step 1: Undefinability of half-regular 3XOR}

\begin{definition}
    A 3XOR instance is \emph{$d$-half-regular} if every variable appears in at most $d$ equations.
\end{definition}

\begin{lemma}\label{lem:HalfRegular}
    $\mathrm{GapDHalfRegular3XOR}(1,\frac12 + \delta)$ is $\FPC$-undefinable for any $\delta>0$, $d>0$
\end{lemma}

In order to see this, we can modify the proof of Theorem~\ref{thm:fpc-xor} as given  in~\cite{atserias2019definable} to apply to half-regular 3XOR instances. In order to show this, we need to quickly sketch the proof from~\cite{atserias2019definable} to see how to alter it to get the strenghtened result:

Theorem~\ref{thm:fpc-xor} is obtained as a direct consequence of the following four lemmas. The statements are quoted verbatim from~\cite{atserias2019definable} and we provide some explanation of the terminology below.

\begin{lemma}[\cite{atserias2019definable}, Lemma~2]
  \label{lem:2ofAD}
For every 3XOR instance $I$ and every integer $k \ge 3$, if $I$ is $k$-locally satisfiable, then
\[
G(I) \equiv_{C_k} G(I_0).
\]
\end{lemma}

\begin{lemma}[\cite{atserias2019definable}, Lemma~3]
  \label{lem:3ofAD}
For every 3XOR instance $I$ and every $c, s \in [0,1]$, the following hold:
\begin{enumerate}
    \item If $I$ is $c$-satisfiable, then $G(I)$ is $c$-satisfiable.
    \item If $I$ is not $s$-satisfiable, then $G(I)$ is not $\left(\frac{1}{2} + \frac{s}{2}\right)$-satisfiable.
\end{enumerate}
\end{lemma}

\begin{lemma}[\cite{atserias2019definable}, Lemma~4]
  \label{lem:4ofAD}
For every two reals $\varepsilon > 0$ and $\delta > 0$ there exists an integer $c > 0$ such that
for every sufficiently large integer $n$ and every matrix 
\[
A \in \{0,1\}^{m \times n},
\]
 where $m = cn$ and each row of $A$ has exactly three ones, if $b$ is chosen uniformly at random in $\{0,1\}^m$,
then, with probability at least $1 - \delta$, both 3XOR instances $Ax = b$ and $G(Ax = b)$
are at most $\left(\frac{1}{2} + \varepsilon\right)$-satisfiable.
\end{lemma}

Note that a 3XOR instance can be expressed as the equation $Ax = b$, where $A$ is a matrix with every row representing an equation, every column representing a variable, and $A[i,j]=1$ if, and only if variable $j$ appears in equation $i$, while $b$ is the vector of equation right-hand sides.

\begin{lemma}[\cite{atserias2019definable}, Lemma~5]
  \label{lemma-from-dawar-atserias}
For every integer $r > 0$ there is a real $\gamma > 0$ such that for every sufficiently
large integer $n$ there is a matrix 
\[
A \in \{0,1\}^{m \times n},
\]
where $m = rn$, such that each row of $A$ has exactly three ones and, for every vector
$b \in \{0,1\}^m$, the 3XOR instance $Ax = b$ is $k$-locally satisfiable for
\[
k \le \gamma n.
\]
\end{lemma}

In order to understand these lemmas we need to define some terminology.  First, we define the map $G$ referred to in Lemmas~\ref{lem:2ofAD} and~\ref{lem:3ofAD}.   It takes a 3XOR instance $I$ and replaces every equation $x+y+z=b$ with 8 equations of the form $x_i+y_j+z_k=b+i+j+k$ for all possible triplets $(i,j,k) \in \{0,1\}^3$. By construction, it is easy to see that $G$ maps any half-regular 3XOR instance to another half-regular instance (any $x_i$ appears in $4$ times as many equations in $G(I)$ as the number of equations $x$ appears in $I$). Given a 3XOR instance $I: Ax=b$, $I_0$ is defined as $Ax=0$.  We omit a rigorous definition of the notion of $k$-local satisfiability referred to in Lemmas~\ref{lem:2ofAD} and~\ref{lemma-from-dawar-atserias} as it is not needed.  We just note that the $k$-locally satisfiability of the instance  $Ax = b$ in Lemma~\ref{lemma-from-dawar-atserias} is established directly from the expansion properties of the bipartite graph for which the $A$ is the biadjacency matrix.  We return to these expansion properties below.

Lemmas~\ref{lem:2ofAD},~\ref{lem:3ofAD} and~\ref{lem:4ofAD} hold for all 3XOR instances, including half-regular ones.  
So in order to establish Lemma~\ref{lem:HalfRegular}, it suffices to show that in the proof of Lemma~\ref{lemma-from-dawar-atserias}, the matrix $A$ can be constructed to satisfy the additional condition that every column has the same number of $1$s (i.e.\ each variable appears in $3r$ equations).  Given this, Lemma~\ref{lem:HalfRegular} follows, since we then have a family of instances of the form $G(Ax=b)$ which are no more than $(\frac{1}{2}+ \epsilon)$-satisfiable and a family of instances of the form $G(Ax=0)$ which are satisfiable.  These families are half-regular by construction and not separable by any class of bounded counting width by Lemma~\ref{lem:2ofAD}.

We mentioned above that the $k$-locally satisfiability of the instance $Ax = b$ is established by taking $A$ to be the biadjacency matrix of an arbitrary bipartite unique-neighbour expander graph with parameters $(rn, n, 3, \alpha n, \beta)$.  This means that it is a bipartite graph with $rn$ nodes on the left, $n$ nodes on the right, which is $3$-left-regular, and an $(\alpha n, \beta)$ unique neighbour expander, which we define next.  

\begin{definition}
    We call a bipartite graph an $(X,Y)$ unique-neighbour expander if for all sets of left-nodes $S$ with size at most $X$, the set of unique neighbours, $N(S)\triangleq \{v \mid \exists! u \in S. (u,v) \in E\}$ has size at least $Y|S|$.
\end{definition}

Thus, to complete the proof, we just need to show that there are arbitrarily large such graphs which are also right regular.  Specifically, we need to show that for some fixed $r\in \mathbb{N^+},\alpha,\beta>0$, there exists a bipartite unique neighbour expander that
\begin{itemize}
    \item has $rn$ left nodes and $n$ right nodes;
    \item is $3$-left-regular and $3r$-right-regular; and
    \item is a \emph{unique-neighbour} $(\alpha n, \beta)$-expander.
\end{itemize}

We show this (using $\beta=1$) by a slight variation of the proof of~\cite[Theorem 4.4]{Vadhan12} (case $D=3$):

Suppose we generate a 3-left-regular, $3r$-right-regular bipartite graph with $rn$ left nodes and $n$ right nodes with the following random process
\begin{enumerate}
    \item Start with an empty bipartite graph with $rn$ left nodes and $n$ right nodes. Randomly partition the left nodes into $r$ groups of $n$.
    \item In $3r$ rounds, do the following: take one of the $r$ groups on the left (taking each group exactly $3$ times) and generate a random matching between the group and the nodes on the right, which has no common edge with the previously generated matchings.
    \item The edges of our resulting graph are the union of these $3r$ matchings.
\end{enumerate}

\begin{lemma}
   A graph generated with the random process above is a $(\alpha n, 1)$ unique-neighbour expander with probability at least $\frac12$ for some $\alpha$.
\end{lemma}

\begin{proof}

Fix a subset $S$ of the left vertices of size $K\leq\alpha n$. List the $3K$ neighbours of $S$ (including multiplicites, i.e. if a right node $v$ has $l$ incoming edges from $S$, it is listed $l$ times) in an arbitrary order, naming them $V_1, V_2, ..., V_{3K}$. Also let $U_1, U_2, ... U_{3K}$ be elements of $S$, each element mentioned three times, such that there is an edge $(U_i, V_i)$ for all $i$. The probability of $V_i$ not being a unique neighbour is at most $\sum_{i\neq j} P(V_i=V_j)$. Given $U_i$, $U_j$ and the round in which the corresponding edges have been selected in the random process, we know that for all $i\neq j$ $P(V_i=V_j)$ is

\begin{itemize}
    \item $0$, if $U_i=U_j$, or the nodes $U_i$ and $U_j$ belong to the same group of $n$, and the edges $(U_i, V_i)$ and $(U_j, V_j)$ are selected in the same round (recall that we select matchings in a round, so the edges must be disjoint)
    \item $\frac1n$, if $U_i$ and $U_j$ belong to different groups of $n$ (the edge choices are independent)
    \item $\frac{1}{n-1},$ if  $U_i$ and $U_j$ belong to the same group but the edges are chosen in a different round. To see this, say the edge $(U_i, V_i)$ is selected in round $k$. Then in the same round, an edge $(U_l, V_l)$ is selected such that $U_l=U_j$. Since in every round, we choose a matching, $(V_l \neq V_i)$, $V_j$ must be different from $V_l$, and with no further constraint, it can be any of the remaining $n-1$ right-nodes, with uniform probability.
\end{itemize}

Thus, 

\[ P(V_i \text{ is not a unique neighbour}) \leq \sum_{i\neq j} P(V_i=V_j) \leq \frac{3K-1}{N-1}\leq \frac{3K}{N}\]

Note that the rest of the calculation is identical to \cite[Theorem 4.4]{Vadhan12}, and the explanation from here is an almost verbatim reproduction of the proof from the book, substituting $D=3$:
\begin{itemize}
    \item The probability that $S$ violates the expander property is
    \[
P\bigl[ |N(S)| \leq  K \bigr]
\le
P\bigl[\text{there are at least } 2K \text{ non-unique neighbour in } V_1,\dots,V_{3K}
\]
\[
\le
\binom{3K}{2K}
\left(\frac{3K}{N}\right)^{2K}.
\]

\item
Let $p_K$ be the probability that any set of size $K$ does not expand with $\beta = 1$. Then

\[
p_K
\le
\binom{N}{K}
\binom{KD}{2K}
\left(\frac{KD}{N}\right)^{2K}
\]
\[
\le
\left(\frac{Ne}{K}\right)^K
\left(\frac{KDe}{2K}\right)^{2K}
\left(\frac{KD}{N}\right)^{2K}
=
\left(\frac{e^3 D^4 K}{4N}\right)^K.
\]

\item Noting $K \le \alpha N$, let
$\alpha = \frac{1}{e^3 D^4}$, so $p_K \le 4^{-K}$. Thus,
\[
P\bigl[ G \text{ is not a unique-neighbour } (\alpha N, 1) \text{ expander} \bigr]
\le
\sum_{K=1}^{\lfloor \alpha N \rfloor} 4^{-K}
<
\frac{1}{2}.
\]
\end{itemize}

\end{proof}

This shows that such expander graph exists, completing the proof of the undefinability gap for half--regular 3XOR.

\subsubsection*{Step 2: Undefinability of regular 3XOR}

Recall that "regular" here means bounded-degree and simple, following the reference papers usage.

A half-regular instance can be converted via a reduction into a regular one by ensuring that any two equations share at most one variable.

A half-regular instance $(X, \mathrm{Eq})$ can be converted into a regular one (call it $(X^*, \mathrm{Eq}^*)$) by replacing every equation $e: x+y+z=b$ with three equations (as done in \cite{Minzer18}): $x + y_e + z_e = b ,\quad
    x_e + y + z_e = b ,\quad    x_e + y_e + z = b$,
where $x_e,y_e$, and $z_e$ are new variables only used for these equations.

As shown in~\cite[remarks preceding Theorem 3.3.1, footnote 14, Section 3.3]{Minzer18}, if $X$ is fully satisfiable then so is $X^*$ and if $X$ is no more than $(\frac12+\delta)$-satisfiable, then $X^*$ is at most $\eta$-satisfiable for some $\eta<1$ (for example, taking $\eta=0.9$ suffices).

The reduction can be defined by a first-order interpretation: The universe can be described by
\[\pi^U(x,a_1,a_2,a_3) \equiv [(x=a_1 \vee x=a_2 \vee x=a_3) \wedge (\mathrm{Eq}_0(a_1,a_2,a_3) \vee \mathrm{Eq}_1(a_1,a_2,a_3))] \vee (x=x_1=x_2=x_3)\]
and we can describe the constraints with:
\begin{align*}
\pi^{\mathrm{Eq}_b}&(x,x_1,x_2,x_3,y,y_1,y_2,y_3, z,z_1,z_2,z_3) \equiv \mathrm{Eq}_b(x,y,z) \\
& \wedge \{ \left[ \neg(x=x_1=x_2=x_3) \wedge  \neg(y=y_1=y_2=y_3) \wedge (z=z_1=z_2=z_3)\right] \\ 
&\vee \left[ \neg(x=x_1=x_2=x_3) \wedge  (y=y_1=y_2=y_3) \wedge \neg(z=z_1=z_2=z_3)\right] \\ 
&\vee \left[(x=x_1=x_2=x_3) \wedge  \neg(y=y_1=y_2=y_3) \wedge \neg(z=z_1=z_2=z_3)\right]
\}.
 \end{align*}
 
\subsection{Shuffling variables} \label{exchanging variables}

One issue that arises with the games constructed in the reduction from Section~\ref{sec:reduction} is that we have a fixed alphabet of size $q = 2^l$ and we associate with each vertex $(U,L)$ an arbitrary bijection between the alphabet and the $2^l$ distinct linear functions on the space $L + H_U$ that satisfy the equations in $U$.  The consistency across different vertices is then enforced by the constraint relations.  In order to turn this into a first-order reduction, we want to choose these bijections in a symmetry-preserving fashion.

Let $I$ be our starting instance of $\prob{3XOR}$ and $I_{2:2}^T = \Theta(I)$ be the transitive $2$-to-$2$ game obtained from the first step of the reduction of Section~\ref{sec:reduction}, and let $X$ be the set of variables of $I$.  Let $\rho \in \text{Sym}_X$ be a permutation of $X$.  This permutation has a natural action on other objects constructed from $X$.  In particular, for an equation $e$ of the form $x+y+z =b$, we write $\rho(e)$ for the equation $\rho(x)+\rho(y)+\rho(z) =b$.  When $U$ is a tuple of such equations, we write $\rho(U)$ for the tuple obtained by applying $\rho$ componentwise to each element of the tuple.  Similarly, for other objects obtained by set and tuple constructions from $X$, we apply the permutation $\rho$ to denote the natural induced action without defining it formally.

Furthermore, we also use $\rho$ to denote the invertible linear map on $\FF{2}^X$ obtained by applying $\rho$ to  the basis $(e_x)_{x \in X}$, and extending linearly to all of $\FF{2}^X$.  Thus, in particular, for a subspace $L \subseteq \FF{2}^X$, $\rho(L)$ denotes the image of this space under this map.

The following is now straightforward.
\begin{lemma}[Shuffling Variables 1] \label{exchange-var-1}
    For any permutation $\rho \in \text{Sym}_X$, if $U$ and $\rho(U)$ are both in $\mathcal{U}$, and $(U,L) \in V(I_{2:2}^T)$, then $\rho(U,L) \in V(I_{2:2}^T)$.
  \end{lemma}
  \begin{proof}
    Since $\rho$ maps the basis of $H_U$ formed by the left-hand sides of the equations in $U$ to the corresponding basis of $H_{\rho(U)}$, we have $\rho(H_U) = H_{\rho(U)}$.  By invertibility of $\rho$, a space $L$ is then linearly independent of  $H_U$ if, and only if, $\rho(L)$ is linearly independent of $H_{\rho(U)}$.
  \end{proof}

Now, we want to choose the bijections between our set of $2^l$ labels and the linear functions associated with a vertex $(U,L)$ in such a way that whenever $(U,L)$ and $\rho(U,L)$ are both vertices in $I_{2:2}^T$, then they commute with $\rho$.  For this, fix a \emph{canonical} space $\FF{2}^{3k}$ of dimension $3k$.  For each $U \in \mathcal{U}$, we write $X_U \subseteq X$ for the set of variables that appear in $U$.  Since $U$ is a sequence of $k$ equations with pairwise disjoint sets of variables, we can fix a bijection between $X_U$ and $[3k]$ which induces an isomorphism $\mu_U: \mathbb{F}_2^{X_U} \rightarrow \mathbb{F}_2^{3k}$. We can naturally extend $\mu_U$ to a map between subspaces: for a subspace $L \subseteq \mathbb{F}_2^{X_U}$, $\mu_U(L)=\{\mu_U(v) \mid v \in L\}$. These bijections can be chosen to be $\rho$-invariant (for all $\rho$), that is,
\[\forall S\in\mathbb{F}_2^{X_U}.\quad\mu_{\rho(U)}(\rho(S)) = \mu_U(S).\]
Moreover, if we choose $\mu_U$ so that the variables in each equation in $U$ are mapped to three consecutive integers in $[3k]$, then there is a fixed subspace $H \subseteq \FF{2}^{3k}$ of dimension $k$ such that $\mu_U(H_U)= H$ for all $U$. Similarly, there is a fixed collection $\mathcal{L}$ of $l$-dimensional spaces such that $\mu_U(\mathcal{L}_U)= \mathcal{L}$.  Thus, we can identify the vertices of  $I_{2:2}^T$ uniquely with pairs $(U,L^*)$ where $U \in \mathcal{U}$ and $L^* \in \mathcal{L}$.  This is to be understood as the representation of the vertex $(U, \mu_U^{-1}(L^*))$.

Similarly, for linear functions $f$ over $L \in \mathcal{L_U}$, we can define
\[(\rho(f))(x) = f(\rho^{-1}(x)) : \rho(L) \rightarrow \mathbb{F}_2 \quad \text{and}\]
\[(\mu_U(f))(x) = f(\mu_U^{-1}(x)) : \mu_U(L) \rightarrow \mathbb{F}_2.\]
Then, a linear function $f$ on $L + H_U$ satisfies the equations in $U$ if, and only if, $\mu_U(f)$ satisfies the equations in $\mu_U(U)$.  Hence, we can interpret in a canonical way the label of a node $(U, \mu_U^{-1}(L^*))$ as a linear function with domain $H + L^*$ satisfying the equations in $\mu_U(U)$.

We now show that this can be consistently applied to the constraints of the game.
\begin{lemma}[Shuffling Variables 2]\label{exchange-var-2}
    Suppose $(U,L), (U',L') \in E(I_{2:2}^T)$ and $\rho(U), \rho(U')$ are both  in $\mathcal{U}$. Then
    \begin{itemize}
        \item $(\rho(U,L), \rho(U',L')) \in E(I_{2:2}^T)$
        \item $\Phi ((U,L), (U',L')) = \Phi(\rho(U,L), \rho(U',L'))$
    \end{itemize}
\end{lemma}
\begin{proof}
    By Lemma \ref{exchange-var-1},  $\rho(U,L), \rho(U',L') \in V(I_{2:2}^T)$.
    Also
    \begin{align*}
        \mathrm{dim}(\rho(L) + H_{\rho(U)}+ H_{\rho(U')})=&\mathrm{dim}(\rho(L + H_U + H_{U'})) = \mathrm{dim}(L + H_U + H_{U'}).
    \end{align*}
    The equalities hold because the mapping $\rho$ is an automorphism of $\mathbb{F}_2^X$. The analogous dimensionality property holds with the mapping of subspaces $(L' + H_U + H_{U'})$ and $(L + L'+ H_U + H_{U'})$. Therefore, the dimensionality constraint for drawing edges is invariant under the action of $\rho$. This proves the first bullet point.

    Then if $(f,f')\in \Phi((U,L),(U',L'))$, it means $\mu_U^{-1}(f)$ and $\mu_{U'}^{-1}(f')$ are consistent on the intersection of their domains. Then $\mu_{\rho(U)}^{-1}(f) = \rho(\mu_U^{-1}(f))$ and $\mu_{\rho(U')}^{-1}(f')=\rho(\mu_{U'}^{-1}(f'))$ are consistent too, meaning $(f,f') \in \Phi(\rho(U,L), \rho(U',L'))$.
Hence \(\Phi((U,L),(U',L')) \subseteq \Phi(\rho(U,L), \rho(U',L'))\).
   Applying the same argument to $\rho^{-1}$ yields the other direction.
\end{proof}

\subsection{The reduction to the transitive game}

We now describe how the reduction $\Theta$ from Section~\ref{sec:transitive} can be given as a first-order interpretation.  Fix an appropriate pair of positive integers $k$ and $l$, the reduction parameters. Given a (regular) $\prob{3XOR}$ instance
\( \mathbb{A} = (X, \mathrm{Eq}_0^\mathbb{A}, \mathrm{Eq}_1^{\mathbb{A}})\),
our interpretation maps it to the following (transitive) $2$-to-$2$ game (with alphabet size $2^l$) $\mathbb{B}$.

\noindent
\textbf{Universe}

The universe of $\mathbb{B}$ consists of tuples of elements of $X$ of length $4k + 2^{3k}$.  These tuples can be seen as broken up into three parts.
\begin{itemize}
    \item The first $3k$ elements ($u_{1,1},\dots,u_{k,3}$) are the $3k$ variables in some $U \in \mathcal{U}$.  To define this, we need to say that they are, in order, the collection of variables of a $k$-tuple of equations, that no variable appears more than once, and that when two variables appear in distinct equations, they do not occur together in some other equation in $\str{A}$.
 
    \item The next $k$ elements  $r_1,\dots,r_k$ define the right-hand sides of the $k$ equations in $U$.  To encode these as binary values, we use $r_i = u_{1,1}$ to encode the value $0$ and $r_i = u_{1,2}$ to encode the value $1$.  Since $u_{1,1}$ and $u_{1,2}$ are distinct, this works and can be specified by a first-order formula.
    \item 
      The next $2^{3k}$ elements also encode bits, using the values of $u_{1,1}$ and $u_{1,2}$ as $0$ and $1$.  Think of these as specifying a subset of $\mathbb{F}_2^{3k}$.  We can write a first-order formula that says that this subset is a subspace $L^*$ of dimension $l$ (since $l$ and $k$ are fixed, the formula is simply a big disjunction over all subspaces).  Finally, we can also write a first-order formula that checks that $L^*$ is in $\mathcal{L}$.
\end{itemize}

For completeness, here is the first-order sentence checking all these conditions.

   \begin{align*}
        \pi^U &= \operatorname*{\bigwedge}_{i=1}^{k} \left[\mathrm{Eq}_0(u_{i,1},u_{i,2},u_{i,3}) \wedge r_i = u_{1,1}\right] \vee \left[\mathrm{Eq}_1(u_{i,1},u_{i,2},u_{i,3}) \wedge r_i = u_{1,2}\right] \\
        &\wedge \operatorname*{\bigwedge}_{(a,i) \neq (b,j)} u_{a,i}\neq u_{b,j} \\
        &\wedge \operatorname*{\bigwedge}_{a \neq b, i, j} \neg \left(
    \exists x \operatorname*{\bigvee}_{(\alpha, \beta, \gamma) \in \mathrm{Perm}(u_{a,i}, u_{b,j}, x)} \mathrm{Eq}_0(\alpha, \beta, \gamma) \vee \mathrm{Eq}_1(\alpha, \beta, \gamma)
    \right) \\
    &\wedge \operatorname*{\bigvee}_{L^* \in \mathcal{L}}\left(\operatorname*{\bigwedge}_{i=0}^{2^{3k}-1} b_i = L^*_i\right).
    \end{align*}
    Where $\mathrm{Perm}(x,y,z)$ describes the set of permutations of $x,y,z$.

We can thus, as required, identify the elements of $\str{B}$ with pairs $(U, L)$ which are the vertices of $\Theta(\str{A})$.

\noindent
\textbf{Relations}

Given two vertices $(U, L)$ and $(U', L')$ of $\str{B}$, the type of constraint between them ($1$-to-$1$, $2$-to-$2$ or no constraint at all) only depends on $\mu_U(L), \mu_{U'}(L'), r, r'$ and $I(U, U')$, where $r$,$r'$ are the vectors of the right-hand sides of the equations and 
\[I(U,U') \triangleq   \{((a,i),(b,j))\in (\{1,\dots,k\}\times\{1,2,3\})^2 \mid u_{a,i} = u'_{b,j}\}.\]
If two pairs of vertices agree on all five of these values, there is a permutation $\rho$ of the variables that will take one to the other and then by Lemma~\ref{exchange-var-2}, they must have the same constraint between them.

Note that each of these five parameters can take only a constant number of different values, so for each constraint $C \in \mathcal{C}_1 \cup \mathcal{C}_2$, there is a finite set $S_C$ of $5$-tuples so that $(U, L)$ and $(U', L')$ are connected by a constraint $C$ if, and only if, $(\mu_U(L), \mu_{U'}(L'), r, r', I(U, U')) \in S_C$. The formula $\pi^C$ defining the relation $C$ in $\str{B}$ simply states that the $5$-tuple corresponding to a pair of vertices is in $S_C$.  This translates to a disjunction of a finite number of cases and is clearly $\FO$-definable.  This concludes the reduction to the transitive game.

\subsection{Weight approximation}\label{weight approximation}

The reduction defined in Section~\ref{final-weighted-2-to-2-game} produces an instance $I^w_{2:2}$ with rational weights.  We now describe how we can get an approximation of this with integer weights, where the weights are furthermore bounded by a polynomial in the number of nodes.  This enables us to represent them directly in structures over $\tau_{\text{(w) 2-to-2}_{q}}$ with only a polynomial blowup in the size of instances.

  The vertices of the integer-weighted game are exactly those in the structure $\str{B}$ above.  The main task is to define the weights, by defining a suitable set $C$ of constraints.  Recall that the vertices of $I^w_{2:2}$ are partitioned into cliques $C_1, \dots, C_m$ based on the $1$-to-$1$ constraints.  Suppose $(U_1,L_1) \in C_i$ and $(U_2,L_2) \in C_j$ are two vertices connected by a $2$-to-$2$ constraint.   Then, the weight of the constraint is
\[\sum_{\substack{U,L,L' \\ L,L' \in \mathcal{L}_U \\ \mathrm{dim}(L \cap L') = l-1}} 1_{(U,L) \in C_i \wedge (U,L') 
    \in C_j} \frac{1}{|\mathcal{U}|} \frac{1}{|\{L,L' \in \mathcal{L}_U \mid \mathrm{dim}(L \cap L') = l-1 \}|} \frac{1}{|C_i||C_j|}.\]
Each of the three factors (apart from the indicator variable) describes the probability of a certain choice in the steps of the random process which define the weights.

Of course, $\frac{1}{|\mathcal{U}|}$ is constant for all pairs $(U_1,L_1),  (U_2,L_2)$.  Similarly, $\frac{1}{|\{L,L' \in \mathcal{L}_U \mid \mathrm{dim}(L \cap L') = l-1 \}|}$ is constant by the symmetry argument presented in Section~\ref{exchanging variables}.  Thus, removing them from the expression does not change the relative weights of the constraints.  Also, the clique size only depends on $(U_1, L_1),  (U_2, L_2)$, so the weight expression (without the normalising factors) simplifies to
\begin{equation}\label{eqn:weight}
  \frac{|\{(U,L,L') \mid (U,L) \in C_i, (U,L') \in C_j\}|}{|C_i||C_j|}.
\end{equation}

 One potential way to turn these rational weights into integer weights would be to multiply them with a common denominator.  This is not a viable option since the number of different-sized cliques grows with the size of the input, making the common denominator too large.  However, we have a workaround: instead of these weights, we give an approximation that does not change the soundness parameter significantly but makes the common denominator of the weights small enough (polynomial as a function of the input size) to be definable.

\begin{lemma}
    Given a weighted 2-to-2 game $(G, \Sigma, \Phi, w)$, whose value is at most $\delta$, any game $(G, \Sigma, \Phi, w')$ where $\forall e \in E(G). \frac{1}{\gamma}<\frac{w(e)}{w'(e)}<\gamma$ has value at most $\delta\gamma^2$.
\end{lemma}

\begin{proof}
The total sum of weights $\sum_{e \in E(G)} w'(e) \geq \frac{1}{\gamma}\sum_{e \in E(G)} w(e)$ since $\frac{w(e)}{w'(e)} < \gamma$ for all $e$.  On the other hand, for any colouring $\chi: V(G) \rightarrow \Sigma$, we have $\sum_{e \in S_{\chi}} w'(e) \leq \gamma\sum_{e \in S_{\chi}} w(e)$ and the result follows.
\end{proof}

So, the idea is to approximate clique sizes so that the number of possible denominators is constant and their product grows only polynomially with the input size, while bounding the change with a suitable multiplicative factor $\gamma$.

Fix a vertex $(U,L)$ in a clique $C_i$.  Recall that $(U', L') \in C_i$ if, and only if, there is a one-to-one constraint between $(U, L)$ and $(U', L')$ in $\str{B}$.  First, let us split the equations in $U'$ into two groups: ``useful'' and ``useless'' ones. An equation in $U'$ is useful (for $U$) if it shares at least one variable with $U$ and useless otherwise. Note that the number of useful equations of $(U',L')$ only depends on $U'$, not on $L'$.

Next, we define an equivalence relation $\equiv_U$ on the vertices of the game as follows: $(U_1, L_1) \equiv_U (U_2,L_2)$ iff
\begin{itemize}
    \item $\mu_{U_1}(L_1) = \mu_{U_2}(L_2)$.
    \item $U_1$ and $U_2$ have the same useful equations (for $U$), and these equations are in the same positions within the $k$-tuple.
    \item The right-hand sides of the equations in $U_1$ and $U_2$ are the same.
    \end{itemize}
    It is easily seen that this is, indeed, an equivalence relation.

    Note that the clique $C_i$ is invariant under the equivalence relation $\equiv_U$: each equivalence class is either contained in $C_i$ or disjoint with it, by Lemma~\ref{exchange-var-2} (choosing $\rho$ to be a permutation that fixes the variables of $U$ and any useful equations).

    Now, for any $f$ with $0\leq f \leq k$, we can establish an upper bound on the number of equivalence classes with $f$ useful equations. Recall that any node $(U',L')$ can be uniquely represented by $U'$ and the subspace $\mu_{U'}(L') = L^* \in \mathcal{L}$:
     \begin{itemize}
     \item The number of possible subspaces $L^*\subseteq\mathbb{F}_2^{3k}$ is at most $2^{2^{3k}}$, as that is an upper bound for $|\mathcal{L}|$ (in fact, it is much smaller, but for our purposes, this upper bound suffices).
     \item The number of ways to choose the positions of the useful equations is $\binom{k}{f}\leq 2^k$.
     \item The number of choices for the right-hand sides of the equations is $2^k$.
     \item Since the $\prob{3XOR}$ instance is regular (each variable appears in at most $d$ equations), the number of equations sharing a variable with $U$ is at most $3kd$, so the number of ways of choosing the useful equations is bounded by $(3kd)^{k}$.
     \end{itemize}
     These bounds are all constants, so the number of equivalence classes within the clique, with $f$ useful equations (call it $\nu_{U, L}^f$) is bounded by a constant $\Psi$ for all $f, U, L$.

     The number of elements in an equivalence class with $f$ useful equations is simply the number of ways to set the remaining $k-f$ equations. This can be approximated by $|\mathrm{Eq}|^{k-f}$. Given $f$ useful equations, the probability of a random set of $k-f$ equations having common variables with $U$, the set of useful equations or each other, or making the $k$-tuple invalid by having two variables from different equations which have a common equation in the $\prob{3XOR}$ instance, converges to zero ($O\left(\frac{k^2}{|X|}\right)$) as the instance size grows, due to the regularity condition. By adding all the approximate sizes of the equivalence classes within $C_i$, we can conclude that the approximation
      \[\chi(\bm
       {\nu_{U, L}}) \triangleq \chi(\nu_{U,L}^0, \nu_{U,L}^1, \dots, \nu_{U,L}^{k}) \triangleq \sum_{f=0}^{k} \nu_{U,L}^f |\mathrm{Eq}|^{k-f} \approx |C_i|  \]
     is accurate within an arbitrarily small factor as the input size grows.
     Using this approximation in the weight expression~(\ref{eqn:weight}), we see that
     \(\prod_{\bm{v} \in \{0,\dots,\Psi\}^{k+1}} \chi(\bm{v})^2\)
is a common denominator of all weights. Multiplying all weights by this number, we get the expression
\begin{align}\label{approx-weight-formula}
\begin{split}
    w((U_1,&L_1), (U_2,L_2)) = |\{(U,L,L') \mid (U,L) \in C_i, (U,L') \in C_j\}| \\  &\cdot \prod_{\bm{v} \in \{0,\dots,\Psi\}^{k+1}} \begin{cases} 
    \chi(\bm{v}) & \text{if } \bm{v} \neq \bm{\nu_{(U_1,L_1)}} \\
    1 & \text{if } \bm{v} = \bm{\nu_{(U_1,L_1)}}
\end{cases}
\cdot \prod_{\bm{v} \in \{0,\dots,\Psi\}^{k+1}} \begin{cases} 
    \chi(\bm{v}) & \text{if } \bm{v} \neq \bm{\nu_{(U_2,L_2)}} \\
    1 & \text{if } \bm{v} = \bm{\nu_{(U_2,L_2)}} \\
\end{cases}
\end{split}
\end{align}
As we see next, we can define a reduction in $\FO$ to $2$-to-$2$ games using these approximate weights.

\subsection{Defining the unweighted game}

Finally, we are ready to show that the construction of an unweighted edge consistent $2$-to-$2$ game (with multiple edges to approximate weights as above) can be given by an $\FO$ interpretation.

\noindent
\textbf{Universe}

We need to define the set of vertices, and the set of constraints.  The elements of the universe are tuples of elements of $X$ (the set of variables of the $\prob{3XOR}$ instance $I$) of length $8k + 1 + 2^{3k+1} + Q$, where $Q$ is a parameter we define below.

A vertex $(U, L)$ is coded by the first  $4k+2^{3k}$ elements of this tuple, as before, followed by a sequence of $0$s.  Recall that we code bits $0$ and $1$ by the first and second elements of the tuple.  The first of these $0$s is to be interpreted as an indicator that the tuple is a vertex (it will be $1$ for a constraint), and the rest are padding to make the length of the tuples match.

A constraint $c$ is coded by a tuple where the first $4k+2^{3k}$ elements represent a vertex $(U,L)$, this is followed by a $1$ (i.e.\ a repeat of the second element of the tuple) and then the next $4k+2^{3k}$ elements represent a second vertex $(U',L')$.  The rest of the tuple codes a unique identifier of the constraint, $\mathrm{ID}$.  We construct the interpretation so that for all fixed $(U, L), (U', L')$, there are $w((U, L),(U', L'))$ different identifiers where $w$ is the approximate weight described above.  We show that for this weight function, there is a formula $W$ which defines a set of exactly $w((U, L),(U', L'))$ tuples extending the description of $(U,L)$ and $(U',L')$.
\begin{lemma}\label{lemma-w}
    There exists $Q \in \mathbb{N}^+$ and a first-order formula $W$ which defines a set $T$ of tuples coding pairs  $(U,L), (U',L')$ together with a $Q$-element unique identifier and such that for each fixed $(U,L), (U',L')$, $T$ contains exactly $w((U,L), (U',L'))$ many tuples extending  $(U,L), (U',L')$.
  \end{lemma}
  The proof of this lemma, constructing the formula $W$ is in Section~\ref{app-defining-W} below.

  Thus, we can define the first-order formulas $\mathrm{Node}$ and $\mathrm{Constraint}$ defining the set of vertices and constraints.  For simplicity, we use $U, L, U', L',\mathrm{ID}$ to describe the sub-tuple of variables in their corresponding parts of the $N$-tuple, where $N= 8k + 1 + 2^{3k+1} + Q$.  Thus, the following formula defines exactly those tuples that are nodes.
\[
    \mathrm{Node}(U,L,\mathrm{IsConstraint},U',L', \mathrm{ID})  \equiv  \mathrm{IsConstraint}=0 \wedge \pi^U{(U,L)} \wedge \bigwedge_{x\in (U',L',\mathrm{ID})} x=0.
  \]
The next formula defines exactly those tuples that are constraints.
\begin{align*}
\mathrm{Constraint}&(U,L,\mathrm{IsConstraint}, U', L', \mathrm{ID}) \equiv \\ & \mathrm{IsConstraint}=1 \wedge \pi^U(U,L) \wedge \pi^U(U',L')   \wedge \bigvee_{C \in \mathcal{C}_2}\pi^C((U,L),(U',L')) \wedge W((U,L),(U',L'), \mathrm{ID}).
\end{align*}

\noindent
\textbf{Constraints}

For each $C_{\pi_1,\pi_2} \in \mathcal{C}_2$, we can construct a formula that defines the set of triples $(x,y,c)$ where $x=(U, L, 0,\dots,0)$ $y=(U', L',0\dots,0)$ and $c=(U, L, 1, U', L', \mathrm{ID})$, such that there is a constraint of type $C$ between $x$ and $y$ and $\mathrm{ID}$ is a valid id of a constraint between them.
\[\Phi_{\pi_1,\pi_2}(x,y,c) \equiv \pi^{C_{\pi_1, \pi_2}}(x,y) \wedge (U,L)=(U_1,L_1) \wedge (U',L') = (U_2,L_2).\]
This completes the proof of Theorem~\ref{FPC-2-to-2}.

\subsection{Defining W}\label{app-defining-W}

To prove Lemma~\ref{lemma-w} we define a first-order formula $W(x,y,z)$ in the vocabulary $\tau_{\prob{3XOR}}$, where $x$, $y$ and $z$ are tuples of free variables.  The formula is such that if $x$ and $y$ are interepreted by the elements coding the nodes $(U,L)$ and $(U',L')$ respectively, then there are exactly $w((U,L)(U',L'))$ assignments of values to the tuple $z$ that make $W$ true.  Here $w(U,L)(U',L')$ is the expression given in Equation~\ref{approx-weight-formula}.

To define $W$, we construct formulas defining various elements of Equation~\ref{approx-weight-formula}.  More precisely, for various numerical expressions $e(x,y)$, which depend on the values assigned to $x$ and $y$, we construct formulas we denote $w_{q,e}(x,y,z)$, where $q$ is the length of the tuple of variables $z$.  These formulas have the property that when $x$ and $y$ are interepreted by the elements coding the nodes $(U,L)$ and $(U',L')$ the number of $q$-tuples that can be assigned to $z$ to make $\omega_{q,e}$ true is exactly $e(x,y)$.    As before, we use $0$ and $1$ to denote the first and second elements of the tuple.   Also, for a first-order formula $\phi(x,y)$, let $1_{\phi}$ denote the indicator variable that $\phi$ is true (under an assignment of values to $x$ and $y$).  

\noindent
\textbf{$e = 1$:}  
\(\omega_{1,e}(x,y,z) \equiv (z=0).\)\\
\textbf{$e = 1_{\phi}$:}
\(\omega_{1,e}(x,y,z) \equiv (z=0) \wedge \phi(x,y).\)

\noindent
\textbf{$e = e_1 \times e_2$:}\\
Given  $\omega_ {q_1,e_1}$ and $\omega_{q_2,e_2}$, we can define
\[\omega_{q_1 + q_2,e}(x,y,z_1,\dots,z_{q_1}, z_{q_1+1}, \dots, z_{q_2}) \equiv \omega_{q_1,e_1}(z_1,\dots,z_{q_1}) \wedge \omega_{q_2,e_2}(z_{q_1+1},\dots,z_{q_2}).\]

\noindent
\textbf{$e = e_1 + e_2$:}\\
Given  $\omega_ {q_1,e_1}$ and $\omega_{q_2,e_2}$, (assuming without loss of generality that $q_2 \geq q_1$, we can define
\begin{align*}
    \omega_{1 + q_2, e}(x,y, z_1, z_2,\dots,  z_{q_2+1}) \equiv &\left[z_1=0 \bigwedge \omega_{q_1,e_1}(z_2, \dots, z_{q_1+1}) \bigwedge \operatorname*{\bigwedge}_{i=q_1+2}^{q_2+1} z_i=0 \right] \\
    &\bigvee\left[z_1=1 \bigwedge \omega_{q_2,e_2}(z_2, \dots, z_{q_2+1}). \right]
\end{align*}

\noindent
\textbf{$e=|\mathbf{Eq}|$:} \\It suffices to take a formula defining the disjoint union of the relations $\mathrm{Eq}_0$ and $\mathrm{Eq}_1$.
\[\omega_{4,e}(x,y,z_1,z_2,z_3,z_4) \equiv \left( z_1 = 0 \land \mathrm{Eq}_0(z_2,z_3,z_4) \right) \vee \left( z_1 = 1 \land \mathrm{Eq}_1(z_2,z_3,z_4)\right). \]

  \noindent
 \textbf{$e = |\{(U_1,L_1,L_2) \mid (U_1,L_1) \in C_i, (U_1,L_2) \in C_j\}|$:}\\
  The numerator in Equation~\ref{eqn:weight} (and a term in Equation~\ref{approx-weight-formula}) is \\$e=  |\{(U_1,L_1,L_2) \mid (U_1,L_1) \in C_i, (U_1,L_2) \in C_j\}|$.  We can get a formula for this by defining exactly this set of tuples.  Here $z$ is a tuple of variables composed of three tuples $z_1$, $z_2$ and $z_3$ where $z_1$ has length $4k$ and each of $z_2$ and $z_3$ is of length $2^{3k}$.
\begin{align*}
\omega_{4k+2*2^{3k},e}(x,y,z) &= \pi^U(z_1,z_2) \wedge \pi^U(z_1,z_3) \wedge \operatorname*{\bigvee}_{C \in \mathcal{C}_2}C((z_1,z_2),(z_1,z_3)  \\
& \wedge \operatorname*{\bigvee}_{C \in \mathcal{C}_1} C((z_1,z_2),x)
\wedge \operatorname*{\bigvee}_{C \in \mathcal{C}_1} C((z_1,z_3),y).
\end{align*}

\noindent
\textbf{Defining the size of the equivalence classes}

Another element of Equation~\ref{approx-weight-formula} are conditions of the form $\nu_ {U, L}^f = r$ for various values of $r$.  We now construct a formula $\nu^{f, \geq r}(x)$ with $4k+2^{3k}$ free variables that expresses the condition $\nu_ {U, L}^f \geq r$ when $x$ is interpreted by the tuple coding $(U,L)$.  In the following, lower case letters $u,l$, possibly with subscript indices always denote tuples of variables of length $4k$ and $2^{3k}$ respectively.
Recall that two elements in the clique are in the equivalence relation $\equiv_{(U, L)}$ if, and only if, their $L$ values are the same and share the same useful equations with the same positions.

We begin by defining some auxiliary formulas.  For any $j \in \{1,\ldots,k\}$, the formula $\mathrm{useful}_j(x,u)$ says of a tuple $u$ that the $j$th equation it represents is useful and the formula $\mathrm{diff}_j(u_1,u_2)$ asserts that the two tuples $u_1$ and $u_2$ differ in the $j$th equation:
\[ \mathrm{useful}_j(x,u) \equiv \bigvee_{i\in \{1,\ldots,3k\}} \left( u_{3(j-1)+1} = x_i \vee u_{3(j-1)+2} = x_i \vee
        u_{3(j-1)+3} = x_i \right); \text{ and} \]
    \begin{align*}
     \mathrm{diff}_j(u_1,u_2) \equiv &  (u_1)_{3(j-1)+1} \neq  (u_2)_{3(j-1)+1} \vee (u_1)_{3(j-1)+2} \neq (u_2)_{3(j-1)+2} \vee \\
      & (u_1)_{3(j-1)+3} \neq (u_2)_{3(j-1)+3} \vee  (u_1)_{3k+j} \neq  (u_2)_{3k+j}.
    \end{align*}

With these, we can define $\nu^{f, \geq r}(x)$ as a formula which asserts the existence of $r$ nodes
\[\exists u_1, l_1, \dots, u_r,l_r \operatorname*{\bigwedge}_i \pi^U(u_i,l_i) ; \]
which are are in the same clique as the node coded by $x$
\[\operatorname*{\bigwedge}_i \operatorname*{\bigvee}_{C \in \mathcal{C}_1} C(x,u_i,l_i) ; \]
all have $f$ useful equations
\[\operatorname*{\bigwedge}_{i \in \{1,\dots,r\}} \left\{\operatorname*{\bigvee}_{S \subseteq \{1,\dots,k\}, |S|=f} \left[\operatorname*{\bigwedge}_{j\in \{1,\dots,k\}} \mathrm{useful}_j(x,u_i) \leftrightarrow j\in S\right]\right\}; \]
and such that no two nodes are $\equiv_{(U,L)}$ equivalent when $x$ is interpreted as $(U,L)$
\[    \operatorname*{\bigwedge}_{i \neq j \in \{1,\dots,r\}} l_i \neq l_j \vee \operatorname*{\bigvee}_{o \in \{1,\dots,k\}} \left( \mathrm{useful}_o(u_i) \land \mathrm{diff}_o(u_i,u_j) \right). \]

Then, as usual, $\nu^{f, r}(x)  \equiv \nu^{f,\geq r}(x) \wedge \neg \nu^{f,\geq (r+1)}(x)$.   To give an expression for $\omega_q(\nu_{U,L}^f)$ for some $q$, we can rewrite it as 
\(\sum_{r=1}^{\Psi} 1_{\nu_ {U,L}^{f,r}} \cdot r\) and construct the expression using the composition rules (constants can be constructed via repeated addition of ones, addition, multiplication and indicator variables are defined above)

\noindent
\textbf{Putting it all together:}

For each term in Equation~\ref{approx-weight-formula}, we have described how to 
define a corresponding formula.  Case splits can be handled via indicator variables and constants by repeatedly adding $1$s.  By a repeated application of the addition and multiplication rules, $W$ can be constructed.

\subsection{Proof of Theorem~\ref{irregular-FPC-2-to-2}}
\label{sec:irregular-proof}

With Theorem~\ref{FPC-2-to-2} in hand, we can now complete the proof of
Theorem~\ref{irregular-FPC-2-to-2}.

Fix $\delta$ with $0 < \delta < 1$ and $j \in \mathbb{N}^{+}$, and let $\gamma = \gamma(\delta)$
be the value determined by Lemma~\ref{reg-to-irregular}. Applying the proof of
Theorem~\ref{FPC-2-to-2} with soundness parameter $\gamma / j^{2}$ yields a $q \in \mathbb{N}^{+}$,
a constant $\eta < 1$, and a first-order reduction which maps satisfiable 3XOR instances to
2-to-2 games of value $1$, and maps instances that are at most $\eta$-satisfiable to 2-to-2
games of value less than $\gamma / j^{2}$. By the contrapositive of
Lemma~\ref{reg-to-irregular}, a game of value less than $\gamma / j^{2}$ satisfies
$\mathrm{Irreg2to2}_{j,q} < \delta$. Hence the same reduction maps at most
$\eta$-satisfiable 3XOR instances to games $I$ with $\mathrm{Irreg2to2}_{j,q}(I) < \delta$,
while satisfiable instances are mapped to games $I$ with $\mathrm{Irreg2to2}_{j,q}(I) = 1$.
Since $\mathrm{Gap3XOR}(1, \eta)$ is not FPC definable, neither is $\mathrm{GapIrreg2to2}_{j,q}(1, \delta)$.

%% file: consequences.tex
\subsection{Unique Games}

An immediate corollary of Theorem~\ref{FPC-2-to-2} is the inapproximability of unique games by any constant factors:

Given a 2-to-2 game $I$, we can map it to a Unique Game $I'$ by splitting every constraint into two: given a constraint of type $C_{\pi_1, \pi_2}$, we can replace it with two 1-to-1 constraints of type $C_{\pi_1}$ and $C_{\pi_2}$.  A colouring of the nodes then satisfies the constraint $C_{\pi_1, \pi_2}$ in $I$ if, and only if, exactly one of the two constraints is satisfied in $I'$. Note that a colouring can only satisfy at most one of the two constraints.  The instance $I'$ is not edge consistent, even when $I$ is.  Nonetheless, this gives a reduction from $\Gap\prob{2-to-2}_q(1,\delta)$ to $\Gap\prob{UG}_q(\frac12, \frac{\delta}{2})$ for any $\delta>0$.

This reduction is clearly FO-definable: the universe consists of the set of vertices of the original game, and for each constraint $c$, we have two constraints $(c,0)$ and $(c,1)$.  Then, for each relation $C_{\pi}$ in the target vocabulary, we can define the set of triples $(x,y,(c,i))$ in it by the formula:
\[\Phi_{\pi}(x,y,(c,i)) \equiv \left( i=0 \land \operatorname*{\bigvee}_{\pi'} \Phi_{\pi,\pi'}(x,y,c) \right) \vee \left( i=1 \land \operatorname*{\bigvee}_{\pi'} \Phi_{\pi',\pi}(x,y,c) \right). \]

From this we can deduce the following. 
\begin{theorem}  \label{FPC-UG2}
    For every $\delta$ with $0< \delta < \frac{1}{2}$, there exists $q \in \mathbb{N}^+$ so 
    that $\Gap\prob{UG}_q(\frac12, \delta)$ is $\FPC$ undefinable.
\end{theorem}
This undefinability gap is stronger in terms of the parameters than the gaps proved by Tucker-Foltz~\cite{TF24}, the only previously known undefinability gaps for Unique Games. However, our construction uses instances with multiple edges labelled with the same constraint relation and hence they are not edge distinct.  Since the gaps in~\cite{TF24} are proved for edge distinct games, they are incomparable to Theorem~\ref{FPC-UG2}.

\subsection{Vertex Cover}\label{sec:vertex-cover}

Another consequence of Theorem~\ref{2-to-2-imperfect}  is the $\NP$-Hardness of approximating the Vertex Cover problem by a factor better than $\sqrt{2}$.  The Unique Games Conjecture implies that nothing better than a factor $2$ approximation is possible.  This is tight since polynomial-time algorithms achieving a $2$ approximation are known and, indeed, $\FPC$ definable~\cite{atserias2019definable} .  Before the results of Khot et al.\ establishing Theorem~\ref{2-to-2-imperfect} the best known inapproximability result, conditional only on $\mathrm{P}\neq\mathrm{NP}$,  was $\approx 1.36$.  Atserias and Dawar~\cite{atserias2019definable} showed a corresponding unconditional $\FPC$ undefinability result.  We improve on this in Theorem~\ref{fpc-is}.

Let $\prob{IS}$ denote the function that gives the size of a maximum independent set in a graph, and $\prob{VC}$ denote the function that gives the size of a minimum vertex cover in a graph, both as a proportion of the total number of vertices.  Since the complement of a vertex cover is an independent set, we have for any graph $G$ that $\prob{IS}(G) = 1 - \prob{VC}(G)$.  We can now state the theorem.

\begin{theorem}[FPC-IS]\label{fpc-is}
    For every $\epsilon, \delta$ with $\delta > 0$ and $0 < \epsilon < 1- \frac{1}{\sqrt2} - \delta$, $\mathrm{GapIS}(1- \frac{1}{\sqrt2} - \delta, \epsilon)$ is not definable in $\FPC$.
\end{theorem}

 This is equivalent to the $\FPC$ undefinability of $\mathrm{GapVertexCover}(1-\epsilon,\frac1{\sqrt{2}} + \delta)$, implying the $\FPC$-inapproximability of vertex cover by a factor smaller than $\sqrt{2}$.

To prove Theorem~\ref{fpc-is}, it suffices to show that the reduction in \cite[Chapter~5]{Minzer18} from  \\$\Gap\prob{Irreg2to2}_{j,q}(1-\epsilon,\delta')$ to $\mathrm{GapIS}(1-\frac{1}{\sqrt{2}} -\delta, \epsilon)$ is definable in $\FO$ for any $\epsilon, \delta$ with a suitable choice of $j,q,\delta'$ such that $\Gap\prob{Irreg2to2}_{j,q}(1-\epsilon,\delta')$ is undefinable (by Theorem~\ref{irregular-FPC-2-to-2}).  We first summarise the reduction and then show its $\FO$-definability.  Note that, though Theorem~\ref{irregular-FPC-2-to-2} is proved with perfect completeness, this does not yield perfect completeness in Theorem~\ref{fpc-is}, since the slack $\delta$ is introduced in the reduction.

We now give a description of the reduction that takes a $2$-to-$2$-game instance $((V, E), \Sigma, \Phi)$ to a node-weighted graph $(V', E', w: V' \rightarrow \mathbb{Q}^+)$ as follows.

Fix a rational $p = \frac{P}{Q} = 1-\frac{1}{\sqrt{2}}-\eta$ for a suitably small $\eta$, depending on $\delta$.  The vertices $V'$ are pairs consisting of a node in $V$ and a subset of $\Sigma$: 
\(V'= V \times \mathcal{P}(\Sigma) \).  
The weights are obtained as a function of the second component:
\(w(v,A) = p^{|A|}(1-p)^{|\Sigma|-|A|}\).  The edge set is defined as follows:
\[E' = \{((u, A_1), (v, A_2) \mid  (A_1 \times A_2) \cap \Phi(e) = \emptyset \text{ for an edge } e \text{ on } u,v\}.\]
We can understand this as giving us a graph whose vertices are nodes of the game, labelled by a set of possible colourings.  The edges ensure that two vertices cannot be in an independent set together if no possbile colourings of the game nodes from the associated set satisfies the corresponding constraint.
 As shown in \cite[Chapter~5]{Minzer18}, this is a reduction from  $\Gap\prob{Irreg2to2}_{j,q}(1-\epsilon',\delta')$ to $\mathrm{GapIS}(1-\frac{1}{\sqrt{2}} -\delta, \epsilon)$.

This reduction can easily be adapted to give unweighted graphs.  First, we can make all the weights integers by multiplying them  by $Q^{|\Sigma|}$.  Thus, let $W(v,A) = Q^{|\Sigma|}w(v,A)$.  Then, we can replace each vertex $(v,A) \in V'$, with weight $w(v,A)$ by a ``cloud'' of $W(v,A)$ vertices.  Vertices in clouds corresponding to $(v,A)$ and $(v',A')$ are adjacent if, and only if, $(v,A)$ and $(v',A')$ are in the weighted version.  Since each cloud contains no edges, any maximal independent set contains either the whole cloud or none of it.  Thus, it is easily seen that the value of $\prob{IS}$ on this unweighted graph is the same as on the weighted original.

We now show that this reduction can be defined as a first-order interpretation.  The dimension of the reduction is $ d= 3 + |\Sigma| + Q^{|\Sigma|}$.  A tuple $t \in V^d$ is to be understood as coding one instance of the pair $(v,A)$.  As before, the first two elements of $t$ are used to code bits $0$ and $1$.  The third component is the vertex $v$.  The next $|\Sigma|$ components are bits coding the set $A$, and the final $Q^{|\Sigma|}$ elements are a binary identifier giving a value up to $W(v,A)$, thus ensuring that we have $W(v,A)$ instances of the vertex $(v,A)$.

We assume (as in Section~\ref{app-defining-W}) we have a formula $W_{\text{VC}}(x)$, where $x$ is a $d$-tuple of variables, such that for any $v_1,v_2,v_3 \in V$ with $v_1 \neq v_2$, and any $|\Sigma|$-tuple $a \in \{v_1,v_2\}^{|\Sigma|}$ coding a set $A$, there are exactly $W(v,A)$ $d$-tuples $t$ extending $v_1,v_2,v_3,a$ such that the last $Q^{|\Sigma|}$ elements of $t$ are all either $v_1$ or $v_2$.

Then, the formula $\delta(x)$ defining the universe of the interpretation is given as 
\[ \delta(x) := x_1 \neq x_2 \land \bigwedge_{i \geq 4} (x_i = x_1 \vee x_i = x_2) \land W_{\text{VC}}(x).\]
This still allows more representatives of a vertex $(v,A)$ then needed, as different choices in the first two components yield different tuples.  To get the universe down to the right size, we take a quotient by the equivalence relation defined by the formula $\epsilon(x,y)$:
\[\epsilon(x,y) :=  x_3 = y_3 \land \bigwedge_{i \geq 4} (x_i = x_1 \Leftrightarrow y_i = y_1).\]
Thus, the set of vertices of the graph are the equivalence classes of the relation defined by $\epsilon$ on the tuples satisfying $\delta$.  Two tuples are equivalent if they have the same third component (and so the same vertex $v$) and the rest of the tuple codes the same binary string.

To define the edge relation, consider any $2$-to-$2$ relation $\Pi \subseteq \Sigma \times \Sigma$.  We construct a first-order formula $\Pi(x,y)$, where $x$ and $y$ are $d$-tuples, that says that if $x$ and $y$ code vertices $(u,A)$ and $(v,B)$ then there are elements $s \in A$ and $t\in B$ such that $(s,t) \in \Pi$.  
\[ \Pi(x,y) := \bigvee_{(i,j) \in \Pi} (x_{i+3} = x_2 \land y_{j+3} = y_2). \]
Given such a formula for each relation $\Pi$, we can define the edge relation in the target graph by the following:
\[ \phi_E(x,y) := \exists z C(z) \land \bigvee_{\Pi} [\Phi_{\Pi}(x_3,y_3,z) \land \neg \Pi(x,y)]. \]

In other words, there exists a constraint $z$, in the general 2-to-2 games universe, that is a $\Pi$-constraint between the underlying 2-to-2 nodes $x_3,y_3$ for some 2-to-2 constraint $\Pi$, with the attached set of labels $(A_x, A_y)$ satisfying the set-property $(A_x \times A_y) \cap \Phi(e) = \emptyset$.

\subsection{Graph Colouring}\label{sec:colouring}

Perhaps the most striking consequence of Theorem~\ref{FPC-2-to-2} is the following.

\begin{theorem}
  \label{colourability}
  For every integer $t \geq  3$,  the class of $3$-colourable graphs is not $\FPC$ separable from those that are not $t$-colourable.
\end{theorem}

Theorem~\ref{colourability} should be contrasted with what is known about the $\NP$-hardness of promise graph colouring.  It is known that it is $\NP$-hard to separate the $3$-colourable graphs from those that are not $5$-colourable~\cite{BBKO21}.  It is conjectured that it is $\NP$-hard to separate the $3$-colourable graphs from those that are not $t$-colourable for all $t \geq  3$, but this is open even for $t=6$.  Thus, Theorem~\ref{colourability} provides the first significant example of an $\FPC$ hardness of approximation result that is open in the classical setting of $\NP$-hardness.

Guruswami and Sandeep~\cite{guruswami2020d} show a reduction from $\Gap\prob{Irreg2to2}_{j,q}(1,\delta)$ with the extra assumption that the constraints are $2 \leftrightarrow 2$ constraints, to the problem of separating $3$-colourable graphs from non-$t$-colourable ones~\cite{DMR09}. We show that this reduction is definable in first-order logic, proving Theorem~\ref{colourability}.  As in the last section, we first summarize the reduction, then show that it can be defined in first-order logic.

The first step reduces $\Gap\prob{Irreg2to2}_{j,q}(1,\delta)$ to $\mathrm{GapColour}(4,t)$ for arbitrary $t \geq 4$ (by choosing a suitable $\delta$). Here, $\mathrm{GapColour}(4,t)$ is the problem of separating $4$-colourable and non-$t$-colourable graphs.

Let $S$ be the set $\{0,1,2,3\} \times \{0,1,2,3\}$ and $T$ be a symmetric, doubly-stochastic, $S \times S$ matrix that additionally satisfies the following two conditions:
\begin{enumerate}
    \item The second largest eigenvalue of $T$ is smaller than $1$.
    \item If $x = (x_1,x_2)$ and $y=(y_1,y_2)$ are elements of $S$, and $\{x_1,x_2\} \cap  \{y_1,y_2\} \neq \emptyset$, then $T(x,y) = 0$.
\end{enumerate}
It is  shown in~\cite[Lemma 10]{guruswami2020d} that such matrices exist.  Having fixed such a $T$, we can define the reduction.  The reduction takes an instance $(V,E,\Sigma,\Phi)$, a $2 \leftrightarrow 2$-game  with $\Sigma = \{0,\ldots,2q-1\}$,  to a graph $(V',E')$.  The nodes in $V'$ are tuples $(v,x_0,\dots,x_{2q-1})$, where $v\in V$ and each $x_i \in \{0,1,2,3\}$.  There is an edge in $E'$ between $(u,x_0,\dots,x_{2q-1})$ and $(v,y_0,\dots, y_{2q-1})$ if there is some constraint $e \in E$ linking $u$ and $v$, such that $\Phi(e) = D_{\pi_1, \pi_2}$ as defined in Section \ref{hourglass-variant}) such that:
\begin{equation}\label{eqn:colour}
    \forall i \in \{0,\dots,q-1\}. T((x_{\pi_1(2i)}, x_{\pi_1(2i + 1)}),(y_{\pi_2(2i)}, y_{\pi_2(2i + 1)}))>0.
\end{equation}
This condition, combined with the second condition on $T$, ensures that given a labeling $\chi: V \rightarrow \Sigma$ that satisfies all constraints, the colouring of $V'$ that assigns $x_{\chi(v)}$ to the node $(v,x_0,\dots, x_{2q-1})$ is a proper $4$-colouring of $(V',E')$.  
On the other hand, if $(V,E,\Sigma,\Phi)$ is not $\delta$-satisfiable, $(V',E')$ does not have an independent set of relative size $\delta'$, for a value of $\delta'$ that tends to $0$ as $\delta$ does. Hence the graph is not $\frac{1}{\delta'}$-colourable. The proof of this can be found in \cite[Section~3.3]{guruswami2020d}.

The second step is a reduction of $\mathrm{GapColour}(4,2^{2^t})$ to $\mathrm{GapColour}(3,t)$ for arbitrary $t\geq3$.  In order to describe this, we first define three graph transformations:
\begin{itemize}
    \item ``$\mathrm{dir}$'', takes an undirected graph $(V,E)$ and gives a directed graph $(V,A)$ where for each edge $\{u,v\} \in E$, we have two directed arcs $(u,v)$ and $(v,u)$ in $A$. 
    \item ``$\mathrm{sym}$'', the opposite of $\mathrm{dir}$: a function that takes a directed graph $(V,A)$ and gives an undirected graph $(V,E)$ where $E = \{ \{u,v\} \mid (u,v) \in A \}$.
    \item ``$\mathrm{arc}$'', the arc graph operator that takes a directed graph $(V,A)$ and gives a directed graph $(A,B)$ whose nodes are the arcs of the original graph and $B$ contains pairs $((a,b),(c,d))$ whenever $b=c$.
\end{itemize}

Note that colourability is not affected by whether the edges are directed or not. The reason for introducing directed graphs is that $\mathrm{arc}$ is defined on directed graphs, and it does not commute with $(\mathrm{dir} \circ \mathrm{sym})$.

We can now introduce the lemmas that guide the reduction:

\begin{lemma}[\cite{Harner_1972}, Theorem 9]
    If a graph $G$ is not $2^t$-colourable, $\mathrm{arc}(G)$ is not $t$-colourable.
\end{lemma}

\begin{lemma}[ \cite{Krokhin2023}, Lemma 4.26]
    Given a $4$-colourable directed graph $G$, $\mathrm{arc}(\mathrm{arc}(G))$ is 3-colourable.
\end{lemma}

This means that the graph transformation $(\mathrm{sym}\circ \mathrm{arc}\circ\mathrm{arc}\circ\mathrm{dir})$ is a reduction from \\$\mathrm{GapColour}(4,2^{2^t})$ to $\mathrm{GapColour}(3,t)$.

Showing the FO-definability of the reduction follows the pattern of the reductions in previous sections.  For the first step, we represent the nodes of $V'$ by tuples $v \in V^{4q + 3}$.  Specifically, we consider such tuples where the first two elements are distinct (i.e.\ $v_1 \neq v_2$), and for each $i$ with $4\leq i \leq 4q$, we have $v_i$ is either $v_1$ or $v_2$.  That is, we represent the values in $\{0,1,2,3\}$ using two bits each.  Thus, the universe and the equality relation are defined by the formulas $\delta(x)$ and $\epsilon(x,y)$ given below.
\[ \delta(x) := x_1 \neq x_2 \land \bigwedge_{i \geq 4} (x_i = x_1 \vee x_i = x_2);\]
\[\epsilon(x,y) :=  x_3 = y_3 \land \bigwedge_{i \geq 4} (x_i = x_1 \Leftrightarrow y_i = y_1).\]

To define the edge relation, as in Section~\ref{sec:vertex-cover}, for each type $\Pi = D_{\pi_1,\pi_2}$ of constraint, we can define a first-order formula $\Pi(x,y)$ which is true of a pair of tuples $u,v \in V^{4q+3}$ if they satisfy the condition in~\ref{eqn:colour}.  This is possible because $T$ is a fixed $16\times 16$ matrix and thus a disjunction over all possibilities will suffice.  With this in hand, the edge relation can again be defined by the formula
\[ \phi_E(x,y) := \exists z C(z) \land \bigvee_{\Pi}[\Phi_{\Pi}(x_3,y_3,z) \land \Pi(x,y)]. \]

Finally, it is straightforward to see that the operations $\mathrm{dir}$, $\mathrm{sym}$ and $\mathrm{arc}$ needed for the second stage of the reduction are definable as first-order interpretations.

%% file: conclusion.tex
We have shown that the reductions involved in the proof of the celebrated proof by Khot, Minzer and Safra of the $2$-to-$2$ games theorem can all be implemented as interpretations in first-order logic.  This means that the $\NP$-hardness they establish of separating nearly satisfiable instances from highly unsatisfiable ones can be turned into an unconditional inseparability result in $\FPC$.  Moreover, the result is achieved with \emph{perfect completeness}: it is impossible to separate with an $\FPC$ sentence the fully satisfiable $2$-to-$2$ games from those that are highly unsatisfiable.

From this result we are able to derive a number of consequences, the most striking of which is that it is impossible to separate with an $\FPC$ sentence the graphs that are $3$-colourable from those that are not $t$-colourable for any constant $t$.  The $\NP$-hardness of such a separation is only conjectured for values $t$ larger than $5$.  Moreover, our result strengthens the undefinabiliety results of Atserias and Dalmau~\cite{AtseriasDalmau22}.  We also obtain strong $\FPC$ undefinability results for approximation of unique games.  In terms of approximation ratios these are an improvement over those of Tucker-Foltz~\cite{TF24}. However, the latter results were obtained for \emph{simple} games while ours are for games with multiple edges.

This work suggests a number of further directions to pursue.  One is an investigation of the $\FPC$ definability of promise constraint satisfaction problems (PCSP).  The $t$-colouring of $3$-colourble graphs is one such example, but PCSP are a very active current area of investigation.  Our results could also be tightened by showing them for simple unweighted instances.  Indeed, we believe that Theorem~\ref{FPC-UG2} could be improved to apply to simple games as well, making it a direct improvement of the results of~\cite{TF24}. For this improvement, it would be sufficient to prove the $\FPC$ analogue of the result of Crescenzi et al. \cite{Crescenzi2001} showing a gap reduction from weighted CSP instances to simple unweighted ones. The proof of Khot, Minzer and Safra applies this reduction to establish Theorem~\ref{2-to-2-imperfect} on simple unweighted games.  This merits further study.

%% file: main.bbl
\begin{thebibliography}{10}

\bibitem{Anderson2017}
Matthew Anderson and Anuj Dawar.
\newblock On symmetric circuits and fixed-point logics.
\newblock {\em Theory of Computing Systems}, 60(3):521–551, July 2017.
\newblock URL: \url{http://dx.doi.org/10.1007/s00224-016-9692-2}, \href
  {https://doi.org/10.1007/s00224-016-9692-2}
  {\path{doi:10.1007/s00224-016-9692-2}}.

\bibitem{AndersonDH15}
Matthew Anderson, Anuj Dawar, and Bjarki Holm.
\newblock Solving linear programs without breaking abstractions.
\newblock {\em J. ACM}, 62:48:1--48:26, 2015.

\bibitem{PCP98.2}
Sanjeev Arora, Carsten Lund, Rajeev Motwani, Madhu Sudan, and Mario Szegedy.
\newblock Proof verification and the hardness of approximation problems.
\newblock {\em J. ACM}, 45(3):501–555, may 1998.
\newblock \href {https://doi.org/10.1145/278298.278306}
  {\path{doi:10.1145/278298.278306}}.

\bibitem{PCP98}
Sanjeev Arora and Shmuel Safra.
\newblock Probabilistic checking of proofs: a new characterization of np.
\newblock {\em J. ACM}, 45(1):70–122, jan 1998.
\newblock \href {https://doi.org/10.1145/273865.273901}
  {\path{doi:10.1145/273865.273901}}.

\bibitem{ATSERIAS20091666}
Albert Atserias, Andrei Bulatov, and Anuj Dawar.
\newblock Affine systems of equations and counting infinitary logic.
\newblock {\em Theoretical Computer Science}, 410(18):1666--1683, 2009.
\newblock Automata, Languages and Programming (ICALP 2007).
\newblock URL:
  \url{https://www.sciencedirect.com/science/article/pii/S0304397508009328},
  \href {https://doi.org/10.1016/j.tcs.2008.12.049}
  {\path{doi:10.1016/j.tcs.2008.12.049}}.

\bibitem{AtseriasDalmau22}
Albert Atserias and V{\'{\i}}ctor Dalmau.
\newblock Promise constraint satisfaction and width.
\newblock In {\em Proceedings of the 2022 {ACM-SIAM} Symposium on Discrete
  Algorithms, {SODA} 2022}, pages 1129--1153. {SIAM}, 2022.
\newblock \href {https://doi.org/10.1137/1.9781611977073.48}
  {\path{doi:10.1137/1.9781611977073.48}}.

\bibitem{atserias2019definable}
Albert Atserias and Anuj Dawar.
\newblock Definable inapproximability: New challenges for {D}uplicator.
\newblock {\em J. Log. Comput.}, 29:1185--1210, 2019.
\newblock \href {https://doi.org/10.1093/LOGCOM/EXZ022}
  {\path{doi:10.1093/LOGCOM/EXZ022}}.

\bibitem{BBKO21}
Libor Barto, Jakub Bulín, Andrei Krokhin, and Jakub Opršal.
\newblock Algebraic approach to promise constraint satisfaction.
\newblock {\em Journal of the ACM}, 68(4):1–66, July 2021.
\newblock URL: \url{http://dx.doi.org/10.1145/3457606}, \href
  {https://doi.org/10.1145/3457606} {\path{doi:10.1145/3457606}}.

\bibitem{BartoK}
Libor Barto and Marcin Kozik.
\newblock Constraint satisfaction problems solvable by local consistency
  methods.
\newblock {\em J. ACM}, 61(1), January 2014.
\newblock \href {https://doi.org/10.1145/2556646} {\path{doi:10.1145/2556646}}.

\bibitem{Crescenzi2001}
Pierluigi Crescenzi, Riccardo Silvestri, and Luca Trevisan.
\newblock On weighted vs unweighted versions of combinatorial optimization
  problems.
\newblock {\em Inf. Comput.}, 167(1):10–26, may 2001.
\newblock \href {https://doi.org/10.1006/inco.2000.3011}
  {\path{doi:10.1006/inco.2000.3011}}.

\bibitem{Dawar15}
Anuj Dawar.
\newblock The nature and power of fixed-point logic with counting.
\newblock {\em ACM SIGLOG News}, 2(1):8–21, jan 2015.
\newblock \href {https://doi.org/10.1145/2728816.2728820}
  {\path{doi:10.1145/2728816.2728820}}.

\bibitem{DawarWangCSL}
Anuj Dawar and Pengming Wang.
\newblock A definability dichotomy for finite valued {CSPs}.
\newblock In {\em 24th {EACSL} Annual Conference on Computer Science Logic,
  {CSL} 2015}, pages 60--77, 2015.

\bibitem{Khot2016}
Irit Dinur, Subhash Khot, Guy Kindler, Dor Minzer, and Muli Safra.
\newblock Towards a proof of the 2-to-1 games conjecture?
\newblock In {\em Proceedings of the 50th Annual ACM SIGACT Symposium on Theory
  of Computing}, STOC 2018, page 376–389, New York, NY, USA, 2018.
  Association for Computing Machinery.
\newblock \href {https://doi.org/10.1145/3188745.3188804}
  {\path{doi:10.1145/3188745.3188804}}.

\bibitem{DMR09}
Irit Dinur, Elchanan Mossel, and Oded Regev.
\newblock Conditional hardness for approximate coloring.
\newblock {\em SIAM Journal on Computing}, 39(3):843–873, January 2009.
\newblock URL: \url{http://dx.doi.org/10.1137/07068062X}, \href
  {https://doi.org/10.1137/07068062x} {\path{doi:10.1137/07068062x}}.

\bibitem{DinurS04}
Irit Dinur and Shmuel Safra.
\newblock On the hardness of approximating label-cover.
\newblock {\em Information Processing Letters}, 89(5):247–254, March 2004.
\newblock URL: \url{http://dx.doi.org/10.1016/J.IPL.2003.11.007}, \href
  {https://doi.org/10.1016/j.ipl.2003.11.007}
  {\path{doi:10.1016/j.ipl.2003.11.007}}.

\bibitem{EbbinghausF99}
Heinz-Dieter Ebbinghaus and Jörg Flum.
\newblock {\em Finite Model Theory}.
\newblock Springer, 2nd edition, 1999.

\bibitem{PCP96}
Uriel Feige, Shafi Goldwasser, Laszlo Lov\'{a}sz, Shmuel Safra, and Mario
  Szegedy.
\newblock Interactive proofs and the hardness of approximating cliques.
\newblock {\em J. ACM}, 43(2):268–292, mar 1996.
\newblock \href {https://doi.org/10.1145/226643.226652}
  {\path{doi:10.1145/226643.226652}}.

\bibitem{guruswami2020d}
Venkatesan Guruswami and Sai Sandeep.
\newblock d-to-1 hardness of coloring 3-colorable graphs with o (1) colors.
\newblock In {\em 47th International Colloquium on Automata, Languages, and
  Programming (ICALP 2020)}. Schloss Dagstuhl-Leibniz-Zentrum f{\"u}r
  Informatik, 2020.

\bibitem{Harner_1972}
C.C Harner and R.C Entringer.
\newblock Arc colorings of digraphs.
\newblock {\em Journal of Combinatorial Theory, Series B}, 13(3):219–225,
  December 1972.
\newblock URL: \url{http://dx.doi.org/10.1016/0095-8956(72)90057-3}, \href
  {https://doi.org/10.1016/0095-8956(72)90057-3}
  {\path{doi:10.1016/0095-8956(72)90057-3}}.

\bibitem{Hastad}
Johan H\r{a}stad.
\newblock Some optimal inapproximability results.
\newblock {\em J. ACM}, 48(4):798–859, jul 2001.
\newblock \href {https://doi.org/10.1145/502090.502098}
  {\path{doi:10.1145/502090.502098}}.

\bibitem{KHOT2002}
Subhash Khot.
\newblock On the power of unique 2-prover 1-round games.
\newblock In {\em Proceedings of the Thiry-Fourth Annual ACM Symposium on
  Theory of Computing}, STOC '02, page 767–775, New York, NY, USA, 2002.
  Association for Computing Machinery.
\newblock \href {https://doi.org/10.1145/509907.510017}
  {\path{doi:10.1145/509907.510017}}.

\bibitem{KHOT2010}
Subhash Khot.
\newblock On the unique games conjecture (invited survey).
\newblock In {\em 2010 IEEE 25th Annual Conference on Computational
  Complexity}, pages 99--121, 2010.
\newblock \href {https://doi.org/10.1109/CCC.2010.19}
  {\path{doi:10.1109/CCC.2010.19}}.

\bibitem{Khot16/2}
Subhash Khot, Dor Minzer, and Muli Safra.
\newblock On independent sets, $2$-to-$2$ games and grassmann graphs.
\newblock Technical Report TR16-124, Electronic Colloquium on Computational
  Complexity (ECCC), 2016.

\bibitem{Khot16}
Subhash Khot, Dor Minzer, and Muli Safra.
\newblock On independent sets, $2$-to-$2$ games and grassmann graphs.
\newblock {\em Theory of Computing}, 21(10):1--55, 2025.
\newblock URL: \url{https://theoryofcomputing.org/articles/v021a010}, \href
  {https://doi.org/10.4086/toc.2025.v021a010}
  {\path{doi:10.4086/toc.2025.v021a010}}.

\bibitem{Krokhin2023}
Andrei Krokhin, Jakub Opršal, Marcin Wrochna, and Stanislav Živný.
\newblock Topology and adjunction in promise constraint satisfaction.
\newblock {\em SIAM Journal on Computing}, 52(1):38–79, February 2023.
\newblock URL: \url{http://dx.doi.org/10.1137/20M1378223}, \href
  {https://doi.org/10.1137/20m1378223} {\path{doi:10.1137/20m1378223}}.

\bibitem{Minzer18}
Dor Minzer.
\newblock {\em On Monotonicity Testing and the 2-to-2 Games Conjecture},
  volume~49 of {\em {ACM} {B}ooks}.
\newblock Association for Computing Machinery, New York, NY, USA, 1 edition,
  2022.
\newblock URL: \url{https://doi.org/10.1145/3568031}.

\bibitem{Raghavendra2008}
Prasad Raghavendra.
\newblock Optimal algorithms and inapproximability results for every csp?
\newblock In {\em Proceedings of the Fortieth Annual ACM Symposium on Theory of
  Computing}, STOC '08, page 245–254, New York, NY, USA, 2008. Association
  for Computing Machinery.
\newblock \href {https://doi.org/10.1145/1374376.1374414}
  {\path{doi:10.1145/1374376.1374414}}.

\bibitem{Khot2018}
Khot Subhash, Dor Minzer, and Muli Safra.
\newblock Pseudorandom sets in grassmann graph have near-perfect expansion.
\newblock In {\em 2018 IEEE 59th Annual Symposium on Foundations of Computer
  Science (FOCS)}, pages 592--601, 2018.
\newblock \href {https://doi.org/10.1109/FOCS.2018.00062}
  {\path{doi:10.1109/FOCS.2018.00062}}.

\bibitem{TF24}
Jamie Tucker-Foltz.
\newblock {Inapproximability of Unique Games in Fixed-Point Logic with
  Counting}.
\newblock {\em {Logical Methods in Computer Science}}, {Volume 20, Issue 2},
  April 2024.
\newblock URL: \url{https://lmcs.episciences.org/13380}, \href
  {https://doi.org/10.46298/lmcs-20(2:3)2024}
  {\path{doi:10.46298/lmcs-20(2:3)2024}}.

\bibitem{Vadhan12}
Salil~P. Vadhan.
\newblock Pseudorandomness.
\newblock {\em Foundations and Trends® in Theoretical Computer FhScience},
  7(1–3):1--336, 2012.
\newblock URL: \url{http://dx.doi.org/10.1561/0400000010}, \href
  {https://doi.org/10.1561/0400000010} {\path{doi:10.1561/0400000010}}.

\end{thebibliography}
